\documentclass[preprint,12pt, 3p]{elsarticle}

\usepackage{amssymb}
\usepackage{amsmath}
\usepackage{pgfplots}
\usepackage{adjustbox}
\usepackage{booktabs}

\usepackage{arydshln}
\usepackage{multirow}

\usepackage{xcolor}
\usepackage{tikz}
\usepackage{subcaption}
\usepackage{enumitem}

\definecolor{lightred}{RGB}{255,153,153}
\definecolor{darkred}{RGB}{139,0,0}
\definecolor{lightgreen}{RGB}{144,238,144}
\definecolor{darkgreen}{RGB}{0,100,0}
\definecolor{lightblue}{RGB}{135,206,250}
\definecolor{darkblue}{RGB}{0,0,139}
\definecolor{brown}{RGB}{165,42,42}

\newcommand{\adamya}[1]{{#1}}
\newcommand{\vikasG}[1]{{#1}}
\newcommand{\ramya}[1]{{#1}}
\newcommand{\venkatR}[1]{{#1}}
\newcommand{\shyam}[1]{{#1}}
\newcommand{\VB}[1]{{#1}}

\newcommand{\AS}[1]{{#1}}

\begin{document}
%\listoftables
\begin{frontmatter}

\title{Cross-domain Recommender Systems via Multimodal Domain Adaptation}

    \author[2]{Adamya Shyam\fnref{equal}}
    \ead{ashyam@cs.du.ac.in}
    \author[1]{Ramya Kamani\fnref{equal}}
    \ead{ramyakamani2000@gmail.com}
    \author[1]{Venkateswara Rao Kagita}
    \ead{venkat.kagita@nitw.ac.in}
    \author[2]{Vikas Kumar~\corref{cor1}}
    \ead{vikas@cs.du.ac.in}

    \address[2]{University of Delhi, Delhi, India}
    \address[1]{National Institute of Technology, Warangal, India}
    \cortext[cor1]{Corresponding author}
    \fntext[equal]{These authors contributed equally to this work.}

\begin{abstract}

Collaborative Filtering (CF) has emerged as one of the most prominent implementation strategies for building recommender systems.  The key idea is to exploit the usage patterns of individuals to generate personalized recommendations. CF techniques, especially for newly launched platforms, often face a critical issue known as the data sparsity problem, which greatly limits their performance. Cross-domain CF alleviates the problem of data sparsity by finding a common set of entities (users or items) across the domains, which then act as a conduit for knowledge transfer. Nevertheless, most real-world datasets are collected from different domains, so they often lack information about anchor points or reference information for entity alignment. This paper introduces a domain adaptation technique to align the embeddings of entities across domains.  Our approach first exploits the available textual and visual information to independently learn a multi-view latent representation for each entity in the auxiliary and target domains. The different representations of the entity are then fused to generate the corresponding unified representation. A domain classifier is then trained to learn the embedding for the domain alignment by fixing the unified features as the anchor points. Experiments on \AS{four} publicly available benchmark datasets indicate the effectiveness of our proposed approach.

\end{abstract}

%%Graphical abstract
%\begin{graphicalabstract}
%\includegraphics{grabs}
%\end{graphicalabstract}

\begin{keyword}
Collaborative Filtering \sep Domain Adaptation \sep Embeddings \sep Latent Representation \sep Textual Features \sep Visual Features.
\end{keyword} 

%\MaketitleBox

\end{frontmatter}

%% \linenumbers

%% main text
\section{Introduction}
\label{sec:Introduction}
Recommender systems are receiving a lot of attention from AI researchers these days due to their widespread \vikasG{uses} across numerous online platforms, including e-commerce~\cite{hwangbo2018recommendation}, social media~\cite{guy2010social}, entertainment~\cite{rosa2015music}, healthcare~\cite{katzman2018deepsurv}, travel~\cite{majumder2022merit},  and so forth. Users' interaction with these platforms leads to the generation of vast volumes of data in terms of likes, dislikes, ratings, reviews, etc.\vikasG{, which} helps in understanding the preferences of the user. However, very few users generally give feedback explicitly on the items they consume, resulting in a data sparsity problem. Considering implicit feedback from the users is adequate \vikasG{and a viable alternative} to understand their preferences for specific items in \vikasG{data-scarce domains}. For instance, it can be reckoned that the user is \ramya{inquisitive} in the item if the user frequently visits or buys it.
Implicit feedback data \ramya{(such as purchases and clicks)} is simple to gather, broadly applicable, and frequently employed in real-world applications. \vikasG{These implicit interactions of a user are often taken as positive feedback in many traditional recommendation algorithms leading to a class imbalance problem~\cite{rendle2012bpr}}. Techniques like negative sampling addressed this problem by randomly labeling some of \venkatR{the unobserved interaction as negative,} yet those could be more efficacious. In contrast, explicit feedback is described as an expressive judgment from a user on the product they have purchased, such as ratings, reviews, and so on. \vikasG{Explicit feedback is considered to be more expressive and verbalized opinions given by individuals. For example, a review ``\textit{Titanic's captivating love story, breathtaking visuals, and emotional performances make it an unforgettable romance classic}" for the Titanic movie can be considered more expressive of the user's taste because it provides specific details and evokes emotions. Additionally, the reference to ``love and romance" highlights the user's appreciation for specific genres.} \VB{Exploiting such explicit feedback may enrich user representation, consequently improving the recommendations. However, the scarcity of such explicit preferences makes the recommendation process challenging.}

\AS{Numerous methods have been developed to leverage both explicit and implicit feedback, along with additional side information for expressive recommendation models in data-scarce domains. Among these, matrix factorization-based collaborative filtering (CF) is the most prevalent approach. This method recommends items to a target user by analyzing the preference patterns of similar users~\cite{kumar2017collaborative, kumar2017proximal, salman2016combining}. Matrix factorization (MF) aims to understand users and items in a shared latent space, allowing their interactions to be modeled as similarities in latent representation. Jeunen et al.~\cite{jeunen2020closed} enhanced this approach by utilizing item-tag data alongside the user-item interactions, thereby capturing additional signals from side information. Similarly, in the context of recommending online courses, the use of side information, such as user-skill and course-skill matrices, along with the user-course interaction matrix, is leveraged to enhance the quality of recommendations~\cite{symeonidis2019multi}. While integration of side information has proved to be beneficial in enhancing recommendations, the scarcity of such data for newly launched systems poses a significant challenge. To address this limitation, cross-domain recommender systems (CDRS) have emerged as a promising solution.} 

\AS{CDRS transfer knowledge from a well-established source domain to a data-scarce target domain, thus enhancing recommendations in data-scarce environments. For example, in cross-domain movie recommendations, knowledge from a source domain like \textit{Books},  with rich user-item interactions, can help provide better recommendations in a target domain like \textit{Movies}, where user interaction data may be sparse. Traditional CDRS techniques often rely on significant entity overlaps across domains to facilitate effective knowledge transfer~\cite{zhang2018cross, kang2019semi, liu2022exploiting}. Yet, the necessity for such overlap between source and target domains often limits their real-world applicability~\cite{liu2022collaborative, li2023semantic}. For instance, transferring knowledge from \textit{Electronics} to \textit{Handmade Crafts} is challenging due to minimal entity overlap and misaligned metadata. To mitigate this problem, recent advancements in CDRS focus on exploiting the rating patterns from the source domain. These rating patterns represent a compressed form of knowledge, which are expanded within the target domain to model the interactions efficiently~\cite{cremonesi2014cross,veeramachaneni2019maximum, veeramachaneni2022hinge, veeramachaneni2022transfer}. Though this approach reduces dependency on overlapping entities, the challenge of differing rating scales across different domains limit the practicality of these models.}

\AS{The latent space alignment approaches in CDRS are designed to be scale-invariant, aiming to align the distribution of user and item representations in the latent space across domains~\cite{wang2021low,yuan2019darec,yu2022cross}. By aligning the embeddings, the focus is on transferring knowledge based on the semantic similarity of their learned representations.  Even though the entities may not be identical, their embeddings capture shared characteristics and behaviors that can be mapped into a common vector space. This allows for knowledge transfer between domains based on the underlying patterns and features of the entities, rather than relying on exact entity matches.  However, achieving such alignment is challenging, as the embeddings of these entities in the lower-dimensional space are independent, leading to diverse distributions. Domain adaptation, a subfield of transfer learning, addresses this challenge by aligning the distributions of entity features through simultaneous updates to their embeddings in both domains. For instance, Yu et al.~\cite{yu2020semi} incorporate MF to learn user and item latent features from the implicit data in the dense and sparse domains while incorporating guidance from user and item embeddings derived from review data during the alignment process. In real-world scenarios, the product review information may also be limited for a sparse domain, such as a newly launched e-commerce platform. Further, reviews are sometimes diverse, and the interpretation is domain-dependent.  For example,  a review for a particular \textit{gadget X} in the \textit{technology} domain, ``the gadget X sleek design and powerful performance make it a must-have for tech enthusiasts", and in the \textit{fashion} domain, ``the gadget X is bulky and outdated design makes it an eyesore", presents contrasting opinions. Alignment only based on textual features may sometimes mislead the optimization.}

% \noindent \vikasG{Most of the recommendation domains, for example, movies, fashion, technology, etc., are very rich with visual content. In recent years, the emergence of deep neural network architectures (DNNs) has expedited the research on domain adaptation based on visual features in several areas of machine learning research~\cite{wang2018visual, wang2018deep}. Although, the area of recommender systems remains largely unexplored in this context~\cite{fernando2013unsupervised, wang2019recsys}.} 
\AS{The rise of deep neural network architectures (DNNs) as well as the availability of visual content in the majority of the recommendation domains like movies, fashion, technology, etc. makes it an area worth exploring. DNNs have expedited the research on visual features-based domain adaption methods in several areas of machine learning research~\cite{wang2018visual, wang2018deep}, yet the area of recommender systems remains largely untouched in this context~\cite{fernando2013unsupervised, wang2019recsys}. Taking the cue from recent research, we propose a CDRS approach that leverages reviews, visuals, and implicit preference information available in target and auxiliary domains for the alignment of embedding corresponding to users and items, facilitating efficient knowledge transfer.} We use domain-invariant textual and visual features as anchor points in the latent space to guide the alignment. We utilize user reviews and product images for extracting textual and visual features, respectively. Further, the learned latent factors through MF are fused with the textual and visual features to construct rich concatenated representations for users as well as items. Subsequently, the source embeddings are aligned with those of the target through a domain adaptation process. This process intends to learn a mapping function to capture the association between the representations in the source and target domains. \adamya{In this work, we use a deep multi-layer perceptron network as proposed by Yu et al.~\cite{yu2020semi} to acquire the relationship.}

\begin{figure}[ht!]
\centerline{\includegraphics[width=14cm, height = 9cm]{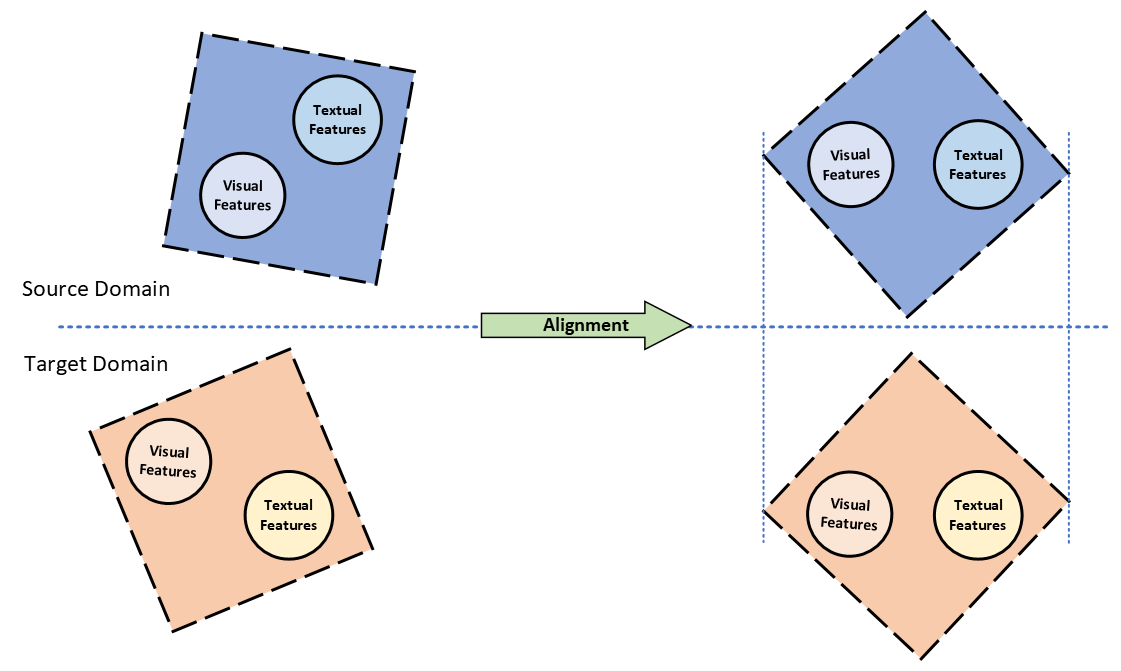}}
\caption{\adamya{Latent factor alignment of user and item in the source and target domain. The distributions of both entities vary in different domains.} %Alignment of user and item distributions in source and target domain. The distributions of user and item are different in different domains.
The textual and visual features are extracted by the same feature extractor models and thus mapped to the same spaces. These extracted features are leveraged in the alignment of embeddings of both domains in the latent space.}
\label{fig:align}
\end{figure}

\AS{The proposed approach can be envisioned as a form of semi-supervised technique. In the source domain, learning involves guidance from both positive and negative samples, whereas in the target domain, learning is exclusively driven by positive examples alongside the domain adaptation technique.} Figure \ref{fig:align} illustrates a graphical abstract view representing the alignment of distributions of users and items in both domains.

%\pagebreak
Our main contributions to the problem are highlighted as follows. 
\begin{itemize}
    \item We first suggest two feature extraction mechanisms, 1) the Textual Feature Extractor, which maps 
    %each user and item into a word semantic space to extract textual features,
    \ramya{user reviews to extract textual features,} and 2) the Visual Feature Extractor maps corresponding product images to extract visual features.
    \item We introduce a Collaborative Filtering (CF) model named Fusion Collaborative Filtering (FCF), in which the fusion of textual and visual features and embeddings are considered for making predictions.  
    \item We introduce a Domain Adaptation Recommendation method, a feature fusion-based method, where two FCF models are trained on source and target domains simultaneously and then connected using an adaptation net. \adamya{The unified representation incorporating textual and visual feature for both users and items serves as the anchor point for guiding the embedding alignment across domains.}%The unified representation of users and items based on textual and visual features serves as anchor points for guiding the embedding alignment across domains. 
    \item We show the efficacy of the proposed models by conducting a \adamya{comprehensive} experimental analysis on four real-world datasets. 
\end{itemize}

The subsequent sections of this paper are organized as follows. An extensive overview of the existing literature and prior relevant studies is presented in Section \ref{sec:Related Work}. Section \ref{sec:Proposed Method} details our proposed approach, including feature extraction approaches and a recommendation framework using domain adaptation. The comparative analysis of our proposed approach with the benchmark methods is outlined in Section \ref{sec:Experiment}, demonstrating the efficacy of our work. Finally, Section \ref{sec:Conclusion} provides the conclusion and discusses potential directions for future research and improvements.

\section{Related Work}
\label{sec:Related Work}

%In this section, we review existing research on the integration of semi-supervised learning with collaborative filtering for single and cross-domain recommendations that make use of auxiliary data such as text and visuals. 

\AS{Recommender systems have long been a focus of research, aiming to provide personalized and relevant recommendations to enhance user experience and aid decision-making~\cite{guo2024dual,xu2021recommendation, isinkaye2015recommendation}. Among the various techniques employed, collaborative filtering (CF) stands out as a prominent approach. CF techniques analyze the behavior and preferences of similar users or items to make predictions~\cite{kant2018leaderrank,hu2018gene,li2020heterogeneousgraphcollaborativefiltering, chen2023heterogeneous}. Latent factor models, a popular variant of CF, uncover latent features representing the underlying characteristics of users and items~\cite{kumar2017collaborative,kumar2017proximal}. Despite their effectiveness, these techniques often face performance limitations due to the sparse nature of the interaction data~\cite{rendle2012bpr,salman2016combining,koren2009matrix}. To address this challenge, combining self-supervised and semi-supervised learning techniques with CF has emerged as a promising solution.}

\AS{Self-supervised learning (SSL) based approaches address data sparsity by generating pseudo-labels from the available data~\cite{ren2024sslrec,wu2021self,yu2023self,xia2022self,xia2023automated}.} Yao et al.~\cite{yao2021self} proposed an architecture-agnostic SSL framework tailored for large-scale neural recommender systems, and introduced a novel data augmentation technique specifically designed for heterogeneous categorical features. \AS{In a similar work, Xia et al.~\cite{xia2022self} proposed a novel self-supervised framework that leverages a hypergraph transformer (SHT) to capture global collaborative relationships among users and items. SHT incorporates a cross-view generative self-supervised learning component to augment user representations and enhance recommendations robustness.} 
In subsequent work, they introduced automated collaborative filtering (AutoCF), a generative self-supervised learning framework with a learnable augmentation paradigm, which facilitates the automated distillation of crucial self-supervised signals~\cite{xia2023automated}. To improve representation discrimination, a masked graph autoencoder is employed to aggregate global information during augmentation by reconstructing the masked subgraph structures.   

\AS{Semi-supervised approaches, in contrast, leverage auxiliary information such as textual and visual features to enrich the representations of users and items~\cite{zhang2014addressing,chen2018semi}. The proposed work builds on the principles of semi-supervised learning, utilizing textual and visual features for cross-domain recommendations (CDR). In the subsequent text, we review the key advancements in semi-supervised methods. Semi-supervised models that incorporate implicit/explicit feedback along with  textual data have demonstrated great potential in addressing the limitations of traditional CF methods~\cite{huang2021semi,yang2017bridging,yu2020semi}.} McAuley et al.~\cite{mcauley2013hidden} enhanced recommendation performance by aligning latent factors derived from ratings with hidden topics extracted from reviews, using topic information as regularization for user and product parameters. While effective, challenges in modeling latent factors and topic extraction can affect overall performance.
\VB{
% In contrast, the semi-supervised approach uses auxiliary information such as textual and visual features to enrich the representation of users and items~\cite{zhang2014addressing, chen2018semi}. 
% The proposed work focuses on the principles of semi-supervised learning, utilizing textual and visual features for cross-domain recommendation. Therefore, we review the existing work on semi-supervised approaches in the subsequent text. Semi-supervised models incorporating implicit/explicit feedback along with textual have emerged as a promising approach to address the limitations of traditional collaborative filtering~\cite{yu2020semi, yang2017bridging,  huang2021semi}. 
Kim et al.~\cite{kim2016convolutional} introduced ConvMF by combining probabilistic matrix factorization with convolutional neural networks by leveraging textual item descriptions. However, ConvMF's reliance on max-pooling limits its ability to fully exploit textual data~\cite{zhang2020content}. Chen et al.~\cite{chen2018neural} proposed NARRE, an attention-based model for generating explainable recommendations by assessing the importance of reviews.
Several other works have also utilized attention networks to leverage the review data for better recommendations~\cite{chen2018neural, xing2019hierarchical,tay2018multi}.  However, noisy or biased reviews can potentially mislead the attention mechanism, limiting the model's ability to generate accurate and reliable recommendations. To overcome these challenges, Wu et al.~\cite{wu2019context} introduced CARL, a context-aware neural model that jointly considers user-item ratings and textual reviews to learn comprehensive user-item representations. }

\AS{Additionally, visual features have also emerged as a valuable source of information for enhancing traditional CF performance.} \VB{
% In addition to textual features, visual features have also emerged as a valuable source of information for enhancing traditional CF performance. 
Zhao et al.~\cite{zhao2016matrix} utilized visual features from posters and still frames for movie recommendations to better understand user preferences.} \AS{Similarly, He et al.~\cite{he2016vbpr} employed deep networks to extract visual features from product images for personalized recommendations. In another approach, Song et al. \cite{song2019gp} proposed GP-BPR, combining multi-modal fashion item data to capture both general aesthetics and personal preferences. Packer et al. \cite{packer2018visually} advanced this field with an explainable clothing recommendation model, utilizing interpretable image representations, while Sagar et al. \cite{sagar2020pai} introduced attribute-wise explanations for personalized outfit recommendations.} \VB{\textit{Though effective, the integration of visual features alone might fail to represent contextual or semantic nuances, leading to less accurate recommendations.}} \AS{Addressing these limitations, recent works focus on combining textual and visual features to improve recommendation accuracy~\cite{liu2020dynamic, hu2023multi, wu2020visual, cheng2019mmalfm}.}
% Recent research also combines textual and visual features to improve recommender systems \cite{liu2020dynamic, hu2023multi, wu2020visual, cheng2019mmalfm}.
\VB{Liu et al. \cite{liu2020dynamic} aligned image regions with text words using topical attention to capture user preferences. The model also considers the contextual influence of the current item to enhance recommendation accuracy, though misalignment between text and images can affect visual representation accuracy. Cheng et al.~\cite{cheng2019mmalfm} introduced a multi-modal aspect-aware latent factor model (MMALFM), incorporating aspect-based analysis to leverage text reviews and item images for capturing user preferences and item features across various aspects.}

\VB{
The aforementioned studies highlight the use of multimodal data in single-domain recommendations. However, real-world applications involve products with limited descriptions or restricted visual data due to privacy concerns~\cite{hu2019transfer, zang2022survey}. Cross-domain recommender systems (CDRS) tackle these issues by transferring knowledge from data-rich source domains to improve recommendations in data-sparse target domains~\cite{kang2019semi, man2017cross, wang2021low, yang2024cross}.} 
\AS{Traditional CDRS methods often rely on substantial entity overlap across domains~\cite{zhao2023cross,zhu2022personalized}. However, these methods fail to perform in scenarios with minimal overlap or misaligned metadata. Recent approaches focus on leveraging the rating patterns from the source domain to enhance recommendations in the target domain~\cite{he2018robust,yuan2019darec,zhang2022cross}. While reducing the dependency on overlapping entities, these methods face challenges with differing rating scales across domains. Domain adaptation, a key technique within this framework, further aids in achieving accurate alignment of user and item distributions by incorporating auxiliary information such as textual and visual data into the adaptation process~\cite{tan2014cross,hu2019transfer}. To facilitate knowledge transfer, feature distribution alignment between domains is achieved through techniques such as adversarial training. The process involves a feature extractor that learns domain-invariant features, while a domain classifier identifies entities unique to each domain. The adversarial optimization ensures that the extractor generates representations that effectively confuse the domain classifier~\cite{ganin2016domain, tzeng2017adversarial, guo2023dan}.} % The approach proposed in \cite{ganin2016domain, tzeng2017adversarial, guo2023dan} aligns feature distributions from both domains to facilitate knowledge transfer. A feature extractor acquires domain-invariant features, while a domain classifier distinguishes entities across domains. Adversarial training optimizes the feature extractor to generate representations that confuse the domain classifier. 
Kanagawa et al.~\cite{kanagawa2019cross} employed a domain adaptation framework integrated with a denoising autoencoder to model item representation using textual features. Jin et al.~\cite{jin2020racrec} filter out the relevant reviews using an adjacency matrix, and then capture the sentiments of those reviews by representing them as vectors, thereby aiding the recommendation task. \AS{Several other studies have utilized textual features for cross-domain recommendations \cite{yu2020semi, wang2019recsys}.} % Works in \cite{yu2020semi, wang2019recsys} also use textual features for a cross-domain recommendation.
\AS{Apart from the textual information, few studies have also utilized visual features to capture the aesthetic preferences, thus enhancing the cross-domain recommendations~\cite{hirakawa2021cross,jaradat2017deep,liu2020exploiting}.} Jaradat~\cite{jaradat2017deep} proposed a deep learning framework for fashion recommendation that extracts visual features from product images and further enhances performance by augmenting textual features. Hirakawa et al.~\cite{hirakawa2021cross} utilizes a convolutional neural network~\cite{jia2014caffe} to extract visual features and incorporate them into a graph neural network to estimate the embedding features. \AS{Liu et al.~\cite{liu2020exploiting} introduced Aesthetic
Cross-Domain Networks (ACDN) to transfer users' domain-independent aesthetic preferences. The visual features are extracted from the product images using a pretrained deep neural architecture, ILGNet~\cite{jin2019ilgnet}, and are incorporated into a cross-domain network. Additionally, cross-connections within the network are employed to facilitate the dual knowledge transfer across domains}. \textit{Previous studies typically employ either textual or visual features in combination with implicit/explicit feedback for domain adaptation.   In contrast, our proposal focuses on improving the representation of entities by integrating both textual and visual features throughout the adaptation process. Textual data captures semantic context and detailed descriptions, while visual data conveys aesthetic and visual appeal, making the recommendations more accurate and robust.}

\section{Proposed Method}
\label{sec:Proposed Method}
\adamya{This section introduces the proposed approach. Initially, we delve into the methods of feature extraction, encompassing textual and visual features extracted from reviews and product images. Subsequently, we elaborate on the approach regarding feature fusion within the collaborative filtering framework, focusing on predicting preferences in both the source and target domains. Finally, we explore our devised semi-supervised domain adaptation method, aimed at aligning embeddings between both domains. This adaptation leverages textual, visual, and implicit feedback data for alignment.} 

%\subsection{Important Notations}
In the subsequent section, $\mathbf{R}$ denotes the implicit interaction matrix, and $\mathbf{\hat{R}}$ is the corresponding dense prediction matrix. The entry $\mathbf{R}_{u i}=1$ denotes that the user $u$ interacted with an item $i$, otherwise $\mathbf{R}_{u i}=0$. $\mathbf{U_{lf}}$, $\mathbf{U_{tf}}$ and $\mathbf{U_{vf}}$ denote the latent factor matrix of \adamya{$M$} users generated from interaction, textual and visual data, respectively. Similarly, $\mathbf{V_{lf}}$, $\mathbf{V_{tf}}$ and $\mathbf{V_{vf}}$ denote the latent factor matrix of \adamya{$N$} items generated from interaction, textual and visual data, respectively. %The total number of users and items are represented by $M$ and $N$, respectively. 
\adamya{\textit{In our work, the superscript $s$ is used for source domain, and likewise, $t$ for target domain.}.} 
% \textit{The superscripts $s$ and $t$ denote source and target domains}.
For example, \adamya{$\mathbf{R^s}$ refers to the interaction matrix in source domain and $\mathbf{R^t}$ refers to that in target domain.} 
Given $\mathbf{R^s}$ and $\mathbf{R^t}$, partially observed interaction matrices, the task is to predict preferences corresponding to unobserved entries in the target domain interaction matrix $\mathbf{R^t}$.

\subsection{Feature Extraction Approaches}

In this section, we discuss the feature extraction process that we have followed for both source and target domains. \textit{Notation without any superscript denotes a general notation for a domain.}

\subsubsection{Textual Feature Extractor} 
\VB{To create textual representations of users and items, we employ the model proposed by Yu et al.~\cite{yu2020semi} over the users' reviews of items. For a specific user, we extract the textual features by utilizing all the reviews provided by the user for different items. Similarly, we utilize all the reviews given to an item by different users to extract its corresponding textual features. We denote the set of the words in all reviews of user $u$ as $R_u$ and similarly $R_i$ for item $i$.}  
% We utilize users' reviews of items to create textual representations for both users and items, employing the model proposed in \cite{yu2020semi}. For a particular user, all reviews given by it to the different items are utilized to extract that user's textual features. Similarly, for textual representation of a specific item, all reviews given to that item by different users are utilized. $R_u$ denotes the set of the words in all reviews considered for user $u$, and similarly $R_i$ for item $i$. 
In order to construct the textual features matrix $\mathbf{U_{tf}}$ and $\mathbf{V_{tf}}$, the model utilizes a semantic word matrix $\mathbf{S}$, a pre-trained matrix trained on GoogleNews corpus using the \textit{word2vec} algorithm \cite{mikolov2013distributed}. Each word $w$ in the matrix is associated with a semantic feature vector $S_w$. The textual features for users and items are constructed using linear combinations of word semantic vectors of the reviews. %We consider users' reviews for items to construct the textual representations of users and items employing the model introduced in \cite{yu2020semi}. To extract the user textual features, the model takes all the reviews the user has given for different items and extracts the numerical features from them. Similarly, for items\ramyaY{,} textual features are extracted from the reviews given by different users. The set of the words in all reviews considered for user $u$ and item $i$ is denoted as $R_u$ and $R_i$, respectively. In order to construct the textual features matrix $\mathbf{U_{tf}}$ and $\mathbf{V_{tf}}$, the model utilizes a semantic word matrix $\mathbf{S}$, a pre-trained matrix trained on GoogleNews corpus using the \textit{word2vec} algorithm \cite{mikolov2013distributed}. Each word $w$ in the matrix is associated with a semantic feature vector $S_w$. To construct textual features for users and items, linear combinations of word semantic vectors of the reviews are used. 
For example, the textual features for user $u$, denoted as $\mathbf{u_{tf}}$, are obtained by summing the semantic word vectors of the words in $R_u$, weighted by the semantic preferences of $u$. \adamya{Mathematically, $\mathbf{u_{tf}} = \Sigma_{w\in{R_u}}a_{uw}S_{w}$, where $a_{uw}$ signifies the weight of word $w$ on the basis of user $u$'s semantic preferences. Similarly, the textual features for item $i$, denoted as $\mathbf{i_{tf}}$, are derived. These weights ($a_{uw}, a_{iv}$) are computed for users and items respectively as in \cite{yu2020semi}. These weights play a crucial role in constructing textual features for users and items, effectively encapsulating their preferences for specific words. These features do not represent sequential information, rather they represent the user and item-specific textual information by extracting important keywords.The textual feature matrices of users and items in the source domain are denoted by $\mathbf{U_{tf}^s} \in \mathbb{R}^{M^s \times K_{1}}$ and $\mathbf{V_{tf}^s} \in \mathbb{R}^{N^s \times K_{1}}$ respectively. Similarly, the same is denoted respectively as $\mathbf{U_{tf}^t} \in \mathbb{R}^{M^t \times K_{1}}$ and $\mathbf{V_{tf}^t} \in \mathbb{R}^{N^t \times K_{1}}$ in target domain.}

\subsubsection{Visual Feature Extractor}
\label{sec:Visual_feature_extractor}
\begin{figure}[!hb]
    \centering
    \includegraphics[width=1.0\textwidth, height=7 cm]{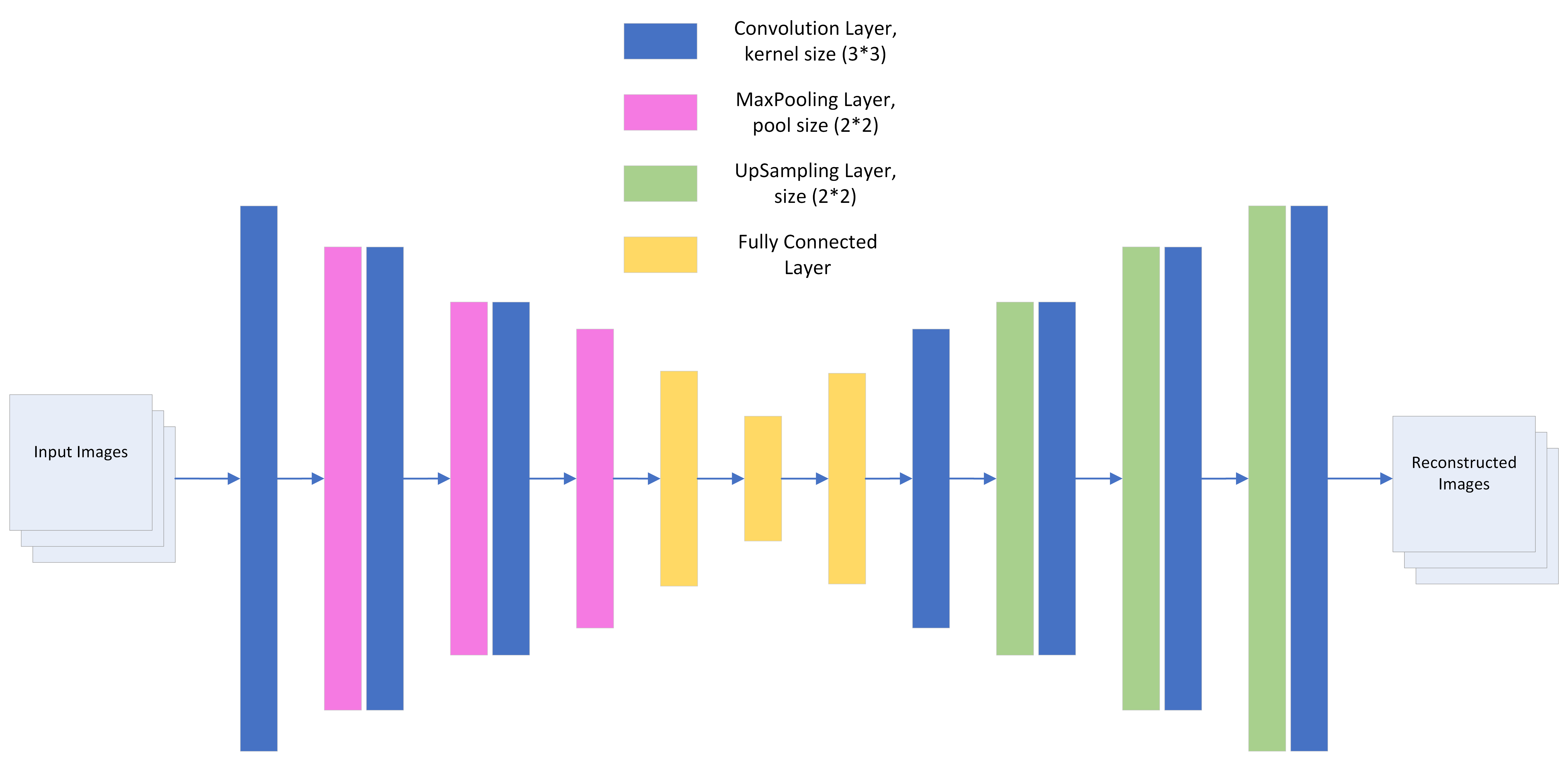}
    \caption{Autoencoder for encoding the images in a lower dimension. The size of an input image is (64, 64, 3), and its corresponding encoded representation is a 300-dimension vector. Different colors indicate different layers in the model. The visual representation for items is taken from the middle layer, which gives an encoded representation.}
    \label{fig:autoencoder}
\end{figure}

\VB{We employ an autoencoder-based architecture to extract the visual features of an item. The autoencoder takes a representative image of the item as input and encodes it into a compact representation. This representation is designed to be concise and adequately informative so that the reconstructed image closely resembles the input image. Hence, we use the intermediate knowledge representation learned by the network as the visual feature vector for the given image.}
% The visual representation of an item is extracted using an autoencoder-based architecture that takes the items' representative image and returns the corresponding latent representation. The autoencoder architecture encodes the given image into a compact representation and then tries to generate the input image from the compact representation. The model ensures that the representation learned by the network is concise and adequately informative to reconstruct the image, which is the same as the input image. Hence, we use this intermediate knowledge representation learned by the network as a visual feature vector of the given image. 
% Figure \ref{fig:autoencoder} illustrates the structure of the proposed autoencoder network. \ramya{The model consists of an encoder network, which compresses the input image into a smaller representation, and a decoder network, which reconstructs the original image from the compressed form.} The encoder uses convolutional and max-pooling layers to extract features and reduce spatial dimensions, followed by fully connected layers to further compress the representation. The decoder uses upsampling and convolutional layers to reconstruct the image. Various experiments have been conducted to determine the number of convolutional and max-pooling layers for this model. 
\VB{The architecture of the autoencoder consists of an encoder and a decoder. The encoder network compresses the input image into a condensed representation, which is further passed to the decoder network for reconstruction of the original image. As illustrated in Figure \ref{fig:autoencoder}, the encoder employs convolutional and max-pooling layers to compress the spatial dimensions and extract features. Subsequently, the fully connected layers further compress the representation. On the other hand, the decoder utilizes upsampling and convolutional layers to reconstruct the image. The optimal number of convolutional and max-pooling layers for this model has been determined by conducting multiple experiments.}

% The chosen autoencoder architecture in Figure \ref{fig:autoencoder} includes three sets of convolutional layers followed by max-pooling layers in the encoder. The decoder consists of three sets of convolutional layers followed by upsampling layers. %This architecture enables accurate image reconstruction, trained using the Adam optimizer to minimize binary cross-entropy loss between the input and the reconstructed output images. 
\adamya{In the chosen autoencoder architecture as shown in Figure \ref{fig:autoencoder}, the encoder comprises three sets of convolutional layers followed by max-pooling layers, and the decoder consists of three sets of convolutional layers followed by upsampling layers.}
This architecture facilitates precise image reconstruction by minimizing the binary cross-entropy loss between the input images and their reconstructed outputs. \adamya{ The optimization process is carried out using the Adam optimizer.} %, which is optimized using the Adam optimizer. 
Further, the datasets used in our experiments contain multiple images of an item provided by different users. We randomly freeze one of the images as the representative image for that item. We do not claim that randomly selected images represent an item best. A strategy that takes into account all different images of an item to construct its latent representation may produce a different result.  Using this autoencoder model, we encode each image into $K_1$ -  dimension vector. We take the size of an input image as $(64, 64, 3)$. Similar to the Textual Feature Extractor model, the visual representation of a user is obtained by taking the mean across the visual representation of items rated by the user. \adamya{The visual feature representation for $u$th user and $i$th item is denoted by $\mathbf{u_{vf}}$ and $\mathbf{i_{vf}}$, respectively. The visual feature matrices in the source domain for users and items are represented as $\mathbf{U_{vf}^s} \in \mathbb{R}^{M^s \times K_{1}}$ and $\mathbf{V_{vf}^s} \in \mathbb{R}^{N^s \times K_{1}}$, respectively. Similarly in target domain, $\mathbf{U_{vf}^t} \in \mathbb{R}^{M^t \times K_{1}}$ represent users' visual features and $\mathbf{V_{vf}^t} \in \mathbb{R}^{N^t \times K_{1}}$ depict items' visual features.}

% We use $\mathbf{u_{vf}}$ and $\mathbf{i_{vf}}$ for the visual representation of $u$th user and $i$th item, respectively. The visual features of users and items in the source and target domains are represented as $\mathbf{U_{vf}^s} \in \mathbb{R}^{M^s \times K_{1}}$, $\mathbf{V_{vf}^s} \in \mathbb{R}^{N^s \times K_{1}}$, $\mathbf{U_{vf}^t} \in \mathbb{R}^{M^t \times K_{1}}$ and $\mathbf{V_{vf}^t} \in \mathbb{R}^{N^t \times K_{1}}$, respectively.

%First, we encode the images of all items which gives the visual representation of $N$ items, $\mathbf{V_{vf}}$, where $\mathbf{i_{vf}}$ represents the visual features for item $i$. Then we obtain the visual representation of each user $u$ by taking the mean of the visual feature vectors of all images of the items that the user $u$ interacted with. For example, if a user $u$ interacted with $n$ items and $I_u$ denotes the set of those $n$ items, then the visual features of $u$ can be obtained as $\mathbf{u_{vf}} = \Sigma_{i \in I_u} \mathbf{i_{vf}} / n $. 

\subsection{Collaborative Filtering}
\label{sec:CF}
In this subsection, we first elaborate on the collaborative filtering framework that we have adopted for our model. We then examine the incorporation of additional knowledge of the features obtained through various feature extraction techniques discussed in the previous sections. These features can be textual or visual, or a fusion of both. 
\adamya{We incorporate these additional features into the collaborative filtering model to train a domain classifier, aiming for the alignment of user and item features, while keeping these embeddings intact. The framework is discussed considering the general representation of user and item latent representations, denoted as $\mathbf{U}$ and $\mathbf{V}$, respectively. However, the specific content within $\mathbf{U}$ and $\mathbf{V}$ may differ across models. The concatenation of two vectors $a$ and $b$ is represented as $[\mathbf{a, b}]$. The prediction of user $u$ preference for item $i$ is formulated as follows.} % We incorporate these additional features into the collaborative filtering model to train a domain classifier for the alignment of user and item embeddings, keeping these features intact. The framework is discussed with the general notion of user and item latent factors denoted as $\mathbf{U}$ and $\mathbf{V}$, respectively. Nonetheless, the content of $\mathbf{U}$ and $\mathbf{V}$ vary from model to model. We use $[\mathbf{a, b}]$ to denote the concatenation of two vectors $a$ and $b$. The preference for a user $u$ and an item $i$ is predicted as follows.  

\begin{equation}
    \hat{\mathbf{R}}_{u i}=\sigma(\mathbf{U}_{u} \mathbf{V}_{i}^T),
\end{equation}
\iffalse
\begin{equation}
    \hat{\mathbf{R}}_{u i}=\sigma([\mathbf{U_{lf}}, \mathbf{U}]_{u}[\mathbf{V_{lf}}, \mathbf{V}]_{i}^T),
\end{equation}
\fi 

\noindent where $\sigma(\cdot)$ denotes sigmoid function. \adamya{Firstly, we train two independent MF-based CF models over the interaction matrices of both the source and target domains, to acquire corresponding users' and items' embeddings. The following cross-entropy loss function is minimized to obtain the optimal parameters of the model}%We first train two independent MF-based CF models over the source and target domains' interaction matrix to learn the embedding for users and items. The optimal parameters of the model are learned through the minimization of the following cross-entropy loss function
\footnote{We use Adam optimizer to learn the optimal parameters that minimize the loss.}.
%We learn this model by minimizing the cross entropy loss function defined in Equation (\ref{eqn:lossCF}) and optimizing using Adam optimizer. In this model, $\mathbf{U_{lf}}$ and $\mathbf{V_{lf}}$ are trainable parameters while $\mathbf{U}$ and $\mathbf{V}$ are fixed. $reg$ indicates the Frobenius norm of $\mathbf{U_{lf}}$ and $\mathbf{V_{lf}}$ and $\lambda$ is the regularization coefficient term.
\begin{equation}
\label{eqn:lossCF}
    \underset{\mathbf{U_{lf}}, \mathbf{V_{lf}}}{min}\; \mathcal{L}=-\sum_{u, i} \mathbf{R}_{u i} \log \hat{\mathbf{R}}_{u i}+(1-\mathbf{R}_{u i}) \log(1-\hat{\mathbf{R}}_{u i})+\lambda (\|\mathbf{U_{lf}}\|_F^2 + \|\mathbf{V_{lf}}\|_F^2)  
\end{equation}
Here, $\lambda >0$ is a regularization parameter.

\noindent \textbf{\emph{Textual Collaborative Filtering (TCF)~\cite{yu2020semi}}} concatenates textual features with latent features and uses the resulting $\mathbf{U}$ and $\mathbf{V}$ to make the prediction. Hence, $\mathbf{U}\equiv[\mathbf{U_{lf}}, \mathbf{U_{tf}}]$ and $\mathbf{V}\equiv[\mathbf{V_{lf}}, \mathbf{V_{tf}}]$ and the task is to learn $\mathbf{U_{lf}}$ and $\mathbf{V_{lf}}$ that minimizes the cross entropy loss mentioned in Equation~\ref{eqn:lossCF}. The approach stimulates the idea of Visual Collaborative Filtering (VCF), which considers visual features akin to textual features in TCF, and Fusion Collaborative Filtering (FCF), which investigates the fusion of textual and visual features in collaborative filtering. We describe these two approaches as follows. 

\noindent\textbf{\emph{Visual Collaborative Filtering (VCF)}} model incorporates the user and item visual features extracted through the autoencoder model in Section~\ref{sec:Visual_feature_extractor} to the user and item latent factors, respectively. The transformed $\mathbf{U}$ and $\mathbf{V}$ are given as follows. 
\begin{equation}
    \mathbf{U} \equiv \mathbf{[U_{lf},U_{vf}]} , \mathbf{V}  \equiv \mathbf{[V_{lf},V_{vf}]}
\end{equation}
\noindent We solve the collaborative filtering optimization problem given in Equation~\ref{eqn:lossCF} and obtain optimal $\mathbf{U_{lf}}$ and $\mathbf{V_{lf}}$. 
%In this subsection, We introduce two novel models, Visual Collaborative Filtering (VCF) model and Fusion Collaborative Filtering (FCF) model. One model is based only on visual features and the other model is based on both textual and visual features which actually takes fusion of these features. The respective features of the model are taken into the CF model and we train the user and item embeddings for that model.
%These embeddings are concatenated with the extracted features according to the model. As such in VCF model, $\mathbf{U}$ and $\mathbf{V}$ represent the visual features of users and items respectively,
%$$
%\mathbf{U} \equiv \mathbf{U_{vf}} , \mathbf{V} \sim \mathbf{V_{vf}}
%$$

%\noindent\textbf{\emph{Fusion Collaborative Filtering (FCF)}} takes the fusion of While in FCF model, these represent the fusion of textual and visual features of users and items respectively. We represent the fusion of features as $f( )$, as such $f(\mathbf{U_{tf}}, \mathbf{U_{vf}})$ represents the fusion of textual and visual features of users and $f(\mathbf{V_{tf}}, \mathbf{V_{vf}})$ is for items. The approach followed for fusion of the features is explained in subsection \ref{sec:Experimental-setting}
\noindent\textbf{\emph{Fusion Collaborative Filtering (FCF)}} blends textual and visual features through the fusion process, and the resulting fusion features will be concatenated with the latent factors. Thus \textbf{U} and \textbf{V} in FCF are defined as follows. 
\iffalse
 $$\mathbf{U} \sim f(\mathbf{U_{tf}}, \mathbf{U_{vf}}) , \mathbf{V} \sim f(\mathbf{V_{tf}}, \mathbf{V_{vf}})$$ 
\fi
\begin{equation}
    \mathbf{U} \equiv [\mathbf{U_{lf}}, f(\mathbf{U_{tf}}, \mathbf{U_{vf}})] ,\mathbf{V} \equiv [\mathbf{V_{lf}}, f(\mathbf{V_{tf}}, \mathbf{V_{vf}})],
\end{equation}
where $f(a,b) = PCA([a,b], K_1)$\footnote{Principal component analysis with a number of principal components as $300$. } that concatenates $a$ and $b$ and reduces the dimension of the resultant vector to $K_1$ to match with the remaining approaches discussed above for %the purpose of 
a fair comparison. Furthermore, the information conveyed by textual and visual representations often overlaps in many cases, motivating the need for dimensionality reduction over the concatenated features. 

\subsection{Domain Adaptation Recommendation}
\VB{As mentioned previously, the embeddings for users and items in both domains are obtained through an independently trained MF-based CF model. For alignment of these embeddings across domains, we employ a domain adaptation process involving a fully connected neural network as a domain classifier. Due to potential differences in the distributions of entities (users and items), we perform separate alignment using the same domain adaptation network model but with different parameters. We leverage the network described by Yu et al.~\cite{yu2020semi} to facilitate the domain classification process. This network has five layers: the input layer size is $K_1 + K_2$, the output layer size is one, and the hidden layers have 64 neurons. Note that the distributions of users and items refer to the concatenated representation of features with embeddings for users and items, respectively.}
% \adamya{As previously mentioned, the users' and items' embeddings for both domains are acquired through an independently trained MF-based CF model. To align these embeddings across domains, we employ a domain adaptation process utilizing a fully connected neural network as a domain classifier. Due to potential differences in the distributions of users and items, we separately align them using the same domain adaptation network model but with different parameters.} We embrace the network given in~\cite{yu2020semi} for the domain classification process. The network has five layers: the input layer is of size $K_1 + K_2$, the output layer size is one, and the size of the hidden layer neurons is 64. Note that the distributions of users and items here indicate the concatenated representation of features with embeddings for users and items, respectively.
%As mentioned earlier, we first learn the embeddings of users and items in both domains using an independently trained MF-based CF model. For aligning the embeddings of users and items across domains, we leverage the domain adaptation process in which we use a fully connected neural network as a domain classifier. Since the distributions of users and items could be different, we align them separately using the same domain adaptation network model with different parameters.
%\venkatR{More details about this network is explained in subsection \ref{sec:Experimental-setting}. I couldn't find the elaboration of the domain classifier network in section 4.2. You have just mentioned that it is a network with 4 layers. }

We represent the domain classifier network as $c(, \Phi)$, where $\Phi$ indicates the parameters of the domain classifier. We use a domain label $d$, representing whether the distributions are from the target or source domains. If the distributions are from the target domain, we take $d=1$; otherwise, $d=0$. We represent the domain label for user $u$ distribution as $d_{u}$ and that of item $i$ as $d_{i}$. The domain classifier takes the distributions from both domains and classifies them to their respective domains. The predicted domain label is denoted as $\hat{d}$.  
%We take a domain classifier represented as $c(, \Phi)$ for domain adaptation process, where $\Phi$ indicates the parameters of domain classifier. We use a domain label $d$, which represents whether the distributions are from target domain or source domain. If the distributions are from target domain, then $d=1$, otherwise $d=0$. The domain classifier takes the distributions from both domains and classifies them to their respective domains and the prediction is represented by a domain label, $\hat{d}$. 
We represent the domain classifier of users and items as $c(, \Phi^U)$ and $c(, \Phi^V)$, respectively, wherein $\Phi^U/\Phi^V$ refers the user/item-distribution domain classifier network parameters. Prediction of domain label for user $u$ distribution is given as $\hat{d_u} = c(, \Phi^U)$. Similarly, for item $i$ distribution, predicted domain label is given as $\hat{d_i} = c(, \Phi^V)$.
The domain classifier networks learn the parameters $\Phi^U$ and $\Phi^V$ to minimize the deviation between the actual and predicted domain labels. Simultaneously, the model also updates the users embedding $\mathbf{U_{lf}}$ and items embedding $\mathbf{V_{lf}}$ of both source and target domains to maximize the deviation between the actual and predicted domain labels. That means the model attempts to learn the source as well as target latent features,  i.e., $\mathbf{U_{lf}^s, U_{lf}^t, V_{lf}^s}$ and $\mathbf{V_{lf}^t}$ in such a way that the source and target distributions of users (or, items) are indistinguishable. At the same time, they are updated to minimize the loss of preference prediction in respective domains (Equation~\ref{eqn:lossCF}). 
%
%For example, consider the following appraoch for alignment of distributions of users. $c(, \Phi^u)$ denotes the domain classifier for user distributions and $\Phi^u$ indicates the parameters of domain classifier of user. We denote the domain label for user distributions as $d_{u}$ which represents whether the user distribution is from source domain or target domain. The domain classifier gives the predictions of domain label: $\hat{d}_{u}=c([\mathbf{U_{lf}}, \mathbf{U}]_{u}, \Phi^{u})$. The domain classifier is trained to discriminate between the distributions of two domains while the embeddings are trained to puzzle the classifier. This results in the alignment of both distributions which are not seperable. While training, we update $\Phi^u$ to minimize the loss of $c(, \Phi^{u})$ and update $\mathbf{U}_{lf}^{s}$ and $\mathbf{U}_{lf}^{t}$ to maximize it. $\mathbf{U}_{lf}^{s}$ and $\mathbf{U}_{lf}^{t}$ are also updated simultaneously to minimize the loss of preference prediction in respective domains. Similar process is followed for alignment of distributions of items. 
\shyam{We have utilized the loss functions proposed in Yu et al.~\cite{yu2020semi}. The loss functions are formulated as follows.}
% The loss functions used are listed as follows.

%\begin{equation}
% \begin{align}
% \label{eqn:lossfns}
% \underset{\mathbf{U_{lf}^s}, \mathbf{V_{lf}^s}}{min} \mathcal{L}^{s}&=-\sum_{u^{s}, i^{s}} \mathbf{R}_{u^{s} i^{s}}^{s} \log \hat{\mathbf{R}}_{u^{s} i^{s}}^{s}+(1-\mathbf{R}_{u^{s} i^{s}}^{s}) \log (1-\hat{\mathbf{R}}_{u^{s} i^{s}}^{s})+\lambda^{s} (\|\mathbf{U_{lf}^s}\|_F^2 + \|\mathbf{V_{lf}^s}\|_F^2) \label{eqn:lossfns_a} \\
% \underset{ \mathbf{U_{lf}^t}, \mathbf{V_{lf}^t} }{min} \mathcal{L}^{t}&=-\sum_{u^{t}, i^{t}} \mathbf{R}_{u^{t} i^{t}}^{t} \log \hat{\mathbf{R}}_{u^{t} i^{t}}^{t}+\lambda^{t} (\|\mathbf{U_{lf}^t}\|_F^2 + \|\mathbf{V_{lf}^t}\|_F^2) \label{eqn:lossfns_b} \\
% \underset{\Phi^U}{min} \mathcal{L}^{u}&=-\sum_{u^{s}, u^{t}} \log \hat{d}_{u^{t}}+\log \left(1-\hat{d}_{u^{s}}\right) \label{eqn:lossfns_c} \\ 
% \underset{\Phi^V}{min} \mathcal{L}^{i}&=-\sum_{i^{s}, i^{t}} \log \hat{d}_{i^{t}}+\log \left(1-\hat{d}_{i^{s}}\right) \label{eqn:lossfns_d} \\
% \underset{\mathbf{U_{lf}^s}, \mathbf{U_{lf}^t}}{max} \mathcal{L}^{u}&=-\sum_{u^{s}, u^{t}} \log \hat{d}_{u^{t}}+\log \left(1-\hat{d}_{u^{s}}\right) \label{eqn:lossfns_a} \label{eqn:lossfns_e} \\
% \underset{\mathbf{V_{lf}^s}, \mathbf{V_{lf}^t}}{max} \mathcal{L}^{i}&=-\sum_{i^{s}, i^{t}} \log \hat{d}_{i^{t}}+\log \left(1-\hat{d}_{i^{s}}\right) \label{eqn:lossfns_f}
% \end{align}
%\end{equation}

\begin{equation}
\underset{\mathbf{U_{lf}^s}, \mathbf{V_{lf}^s}}{min} \mathcal{L}^{s} = -\sum_{u^{s}, i^{s}} \mathbf{R}_{u^{s} i^{s}}^{s} \log \hat{\mathbf{R}}_{u^{s} i^{s}}^{s}+(1-\mathbf{R}_{u^{s} i^{s}}^{s}) \log (1-\hat{\mathbf{R}}_{u^{s} i^{s}}^{s})+\lambda^{s} (\|\mathbf{U_{lf}^s}\|_F^2 + \|\mathbf{V_{lf}^s}\|_F^2) \label{eqn:lossfns_a}
\end{equation}
\begin{equation}
\underset{ \mathbf{U_{lf}^t}, \mathbf{V_{lf}^t} }{min} \mathcal{L}^{t} = -\sum_{u^{t}, i^{t}} \mathbf{R}_{u^{t} i^{t}}^{t} \log \hat{\mathbf{R}}_{u^{t} i^{t}}^{t}+\lambda^{t} (\|\mathbf{U_{lf}^t}\|_F^2 + \|\mathbf{V_{lf}^t}\|_F^2) 
\label{eqn:lossfns_b}
\end{equation}
\begin{equation}
\underset{\Phi^U}{min} \mathcal{L}^{u} = -\sum_{u^{s}, u^{t}} \log \hat{d}_{u^{t}}+\log \left(1-\hat{d}_{u^{s}}\right) 
\label{eqn:lossfns_c} 
\end{equation}
\begin{equation}
\underset{\Phi^V}{min} \mathcal{L}^{i} = -\sum_{i^{s}, i^{t}} \log \hat{d}_{i^{t}}+\log \left(1-\hat{d}_{i^{s}}\right) 
\label{eqn:lossfns_d}
\end{equation}
\begin{equation}
\underset{\mathbf{U_{lf}^s}, \mathbf{U_{lf}^t}}{max} \mathcal{L}^{u} = -\sum_{u^{s}, u^{t}} \log \hat{d}_{u^{t}}+\log \left(1-\hat{d}_{u^{s}}\right) 
\label{eqn:lossfns_e}
\end{equation}
\begin{equation}
\underset{\mathbf{V_{lf}^s}, \mathbf{V_{lf}^t}}{max} \mathcal{L}^{i} = -\sum_{i^{s}, i^{t}} \log \hat{d}_{i^{t}}+\log \left(1-\hat{d}_{i^{s}}\right) 
\label{eqn:lossfns_f}
\end{equation}

\noindent 
The embeddings $\mathbf{U_{lf}^s}$, $\mathbf{V_{lf}^s}$, $\mathbf{U_{lf}^t}$ and $\mathbf{V_{lf}^t}$ are initially trained in the CF model explained in Section \ref{sec:CF} and further optimized using the optimization problems mentioned in Equations (\ref{eqn:lossfns_a}, \ref{eqn:lossfns_b}, \ref{eqn:lossfns_e}, \ref{eqn:lossfns_f}) leveraging textual and visual features.
%In equation (3), 
$\mathbf{U_{lf}^s}$ and $\mathbf{V_{lf}^s}$ (respectively, $\mathbf{U_{lf}^t}$ and $\mathbf{V_{lf}^t}$) are optimized to minimize the loss of preference prediction in the source domain (respectively, target domain). The parameters $\Phi^U$ and $\Phi^V$ of domain classifiers of users and items are optimized to minimize the loss of the corresponding domain classifier, as mentioned in Equations (\ref{eqn:lossfns_c}) and (\ref{eqn:lossfns_d}). Simultaneously, $\mathbf{U_{lf}^s}$ and $\mathbf{U_{lf}^t}$ (respectively, $\mathbf{V_{lf}^s}$ and $\mathbf{V_{lf}^t}$) are optimized to maximize the loss of domain classifier of the user (respectively, item). The updation of these parameters is performed as follows.
%where $\mathcal{L}^{s}$ and $\mathcal{L}^{t}$ represent the loss functions for preference prediction in the source and target domains, respectively. Similarly, $\mathcal{L}^{u}$ and $\mathcal{L}^{i}$ denote the loss functions for domain classification of users and items, respectively. All these loss functions are cross entropy-based, except for $\mathcal{L}^{t}$ which uses positive samples for supervision in the target domain and incorporates domain adaptation for negative supervision. The preference predictions for the source and target domains are denoted by $\hat{\mathbf{R}}_{u^{s} i^{s}}^{s}$ and $\hat{\mathbf{R}}_{u^{t} i^{t}}^{t}$, respectively, as given in Equation (\ref{eqn:lossfns}). Here, $u^{s}$ and $i^{s}$ refer to a user and item from the source domain, while $u^{t}$ and $i^{t}$ correspond to a user and item from the target domain. The regularization coefficients are denoted by $\lambda^{s}$ and $\lambda^{t}$, and $reg^{s}$ and $reg^{t}$ represent the Frobenius norms of $\left\{\mathbf{U}_{lf}^{s}, \mathbf{V}_{lf}^{s}\right\}$ and $\left\{\mathbf{U}_{lf}^{t}, \mathbf{V}_{lf}^{t}\right\}$, respectively. The corresponding model parameters are updated as follows:
\begin{equation}
    \mathbf{U}_{lf}^{s} \leftarrow \mathbf{U}_{lf}^{s}-\nabla_{\mathbf{U}_{lf}^{s}}(\eta^{s} \mathcal{L}^{s}-\eta^{-} \mathcal{L}^{u}) , 
\end{equation}
\begin{equation}
    \mathbf{V}_{lf}^{s} \leftarrow \mathbf{V}_{lf}^{s}-\nabla_{\mathbf{V}_{lf}^{s}}(\eta^{s} \mathcal{L}^{s}-\eta^{-} \mathcal{L}^{i}) ,
\end{equation}
\begin{equation}
    \mathbf{U}_{lf}^{t} \leftarrow \mathbf{U}_{lf}^{t}-\nabla_{\mathbf{U}_{lf}^{t}}(\eta^{t} \mathcal{L}^{t}-\eta^{-} \mathcal{L}^{u}) , 
\end{equation}
\begin{equation}
    \mathbf{V}_{lf}^{t} \leftarrow \mathbf{V}_{lf}^{t}-\nabla_{\mathbf{V}_{lf}^{t}}(\eta^{t} \mathcal{L}^{t}-\eta^{-} \mathcal{L}^{i}) ,
\end{equation}
\begin{equation}
    \Phi^{u} \leftarrow \Phi^{u}-\eta^{+} \nabla_{\Phi^{u}} \mathcal{L}^{u} ,
\end{equation}
\begin{equation}
    \Phi^{i} \leftarrow \Phi^{i}-\eta^{+} \nabla_{\Phi^{i}} \mathcal{L}^{i} ,
\end{equation}

\noindent  where $\nabla_{\mathbf{X}} f(\mathbf{X})$ is the gradient of $f(\mathbf{X})$ with respect to $\mathbf{X}$ and $\eta^{s}, \eta^{t}, \eta^{+}$and $\eta^{-}$ are learning rates.

\begin{figure*}
    %\adjustbox{max width=\textwidth}{
    \centering
    \includegraphics[height=5.7in]{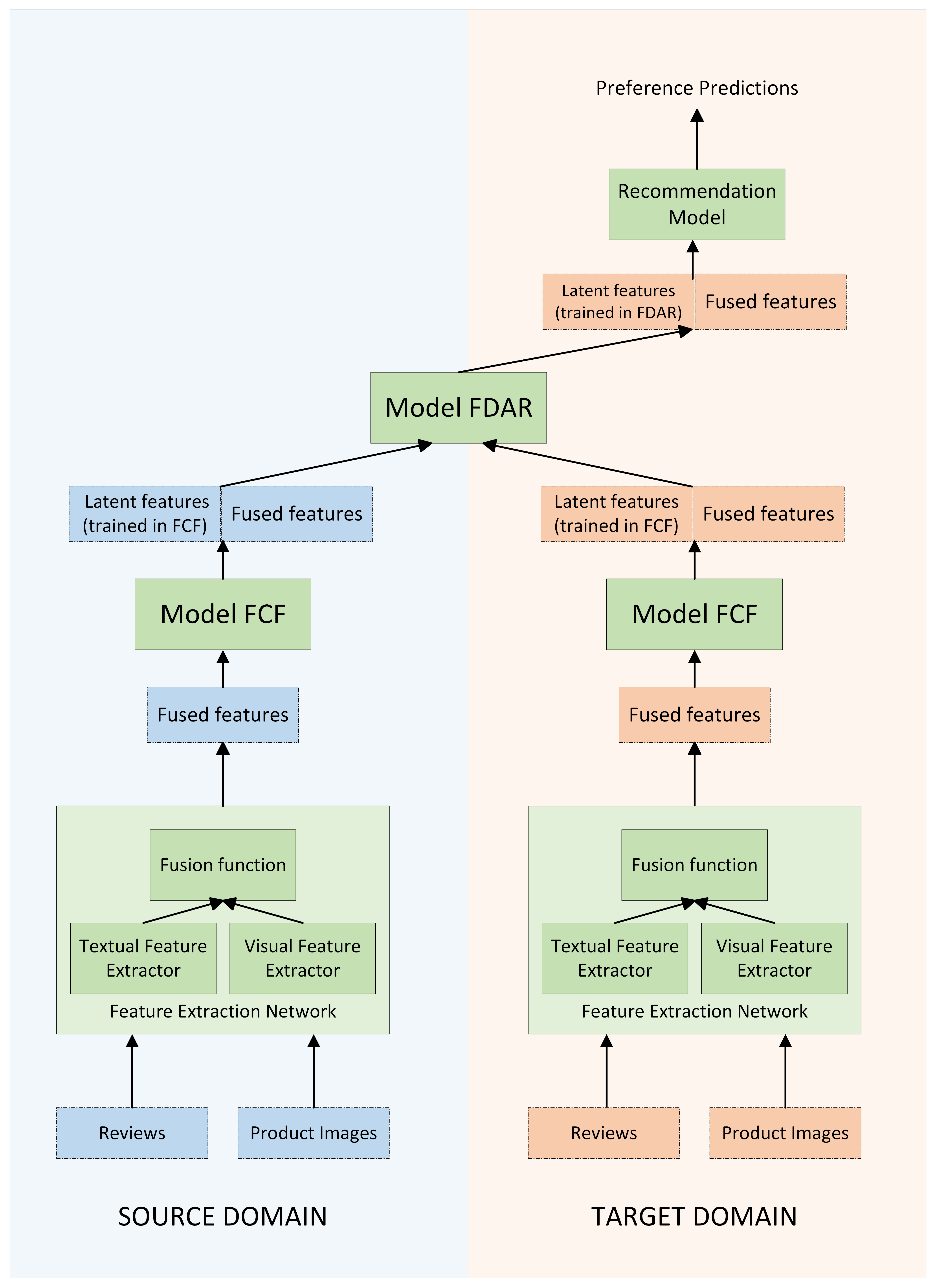} %}
    \caption{Outline of the proposed approach.}
    \label{fig:model}
\end{figure*}

 Figure \ref{fig:model} illustrates the overall working of our proposed model.  In summary, we \adamya{propose} two novel models for cross-domain recommendation: Visual Feature-based Domain Adaptation Recommendation (VDAR) and Feature Fusion-based Domain Adaptation Recommendation (FDAR).  These are cross-domain models \adamya{that exploit the knowledge learned from source domain to perform recommendations in the target domain.} %for recommending items in the target domain using information learned from the source domain.
The basic models used in single domains are VCF and FCF, respectively. Thus in VDAR, the features used are only visual features of both domains, while in FDAR, a fusion of textual and visual features is considered for domain adaptation. We train the two basic CF models of respective domain adaptation recommendation systems, and using the adaptation net, we align the embeddings of users and items. Only positive labels are used for positive supervision of the target domain, considering that negative sampling induces noise in the labels. But supervising only with positive labels creates a bias in predicting all items as positive. 
\adamya{To mitigate this problem, the domain adaptation method performs negative sample supervision on the source domain, enabling transfer of information from source to the target domain.}
% To alleviate this problem, the domain adaptation method also, performs negative sample  supervision on the source domain, enabling the transfer of knowledge acquired from the source domain to the target domain. \ramya{enabling the transfer of knowledge acquired from the source domain to the target domain.} %and the knowledge learned based on the source domain is transferred to the target domain. %Hence, we take the source domain as denser than that of the target domain, and negative sampling is done on the source domain to get the negative samples. 
Thus the approach of domain adaptation used in these models is semi-supervised. We extract textual and visual features from both domains and fix them as pretrained features, which we further use in the domain adaptation process to align embeddings. Experiments show that considering both features as in the FDAR model performs better than considering only visual or textual elements as in VDAR or TDAR \cite{yu2020semi}, respectively.

\section{Experiment}
\label{sec:Experiment}
This section provides a comprehensive overview of the datasets used for the experiments, experimental settings, and benchmark techniques used for comparison and reports the empirical analysis results. 

\subsection{Datasets}
\AS{We have utilized the publicly available Amazon review datasets\footnote{https://cseweb.ucsd.edu/~jmcauley/datasets/amazon\_v2/}, a repository of datasets from a vast category of domains~\cite{ni2019justifying}. For our work, we have considered four categories of datasets, namely \textit{Cell Phones and Accessories}, \textit{Electronics}, \textit{Movies and TV}, and \textit{Video Games} datasets. A summary of the datasets used for our experiments is provided in Table~\ref{tab:stats}. We performed $k$-core filtering (removing all users and items with less than $k$ interactions) over both datasets. For the source dataset, we applied 8-core filtering to increase its density, whereas we chose $k=5$ for the target domain. Further, we ensured that each unique item has at least one corresponding image available in the dataset. 
}

\begin{table}[!ht]
\centering
\renewcommand{\arraystretch}{1.25}
\caption{Statistics of datasets}
\label{tab:stats}
\adjustbox{max width=0.65\linewidth}{
\begin{tabular}{lccccc}
\hline
Dataset & Interaction & User & Item & Sparsity\\
\hline
Cell Phones and Accessories & 36166 & 5514 & 3842 & $99.8293\%$ \\
Electronics & 118607 & 7984 & 5355 & $99.7225 \%$ \\
Movies and TV & 255190 & 16862 & 2786 & $99.4568 \%$ \\
Video Games & 34639 & 4692 & 1299 & $99.4317 \%$ \\

\hline
\end{tabular}
}
\end{table}

\subsection{Baselines}
\adamya{In this section, we discuss the state-of-the-art approaches that we have considered for comparison with our proposed models.} %We compare the performance of our proposed models with the following state-of-the-art approaches.
\begin{itemize}%[noitemsep]
    \item \textbf{Textual Memory Network (TMN)}~\cite{yu2020semi}: This method utilizes domain-invariant features extracted from the textual data to represent users and items.  
    
    \item \textbf{Textual Collaborative Filtering (TCF)}~\cite{yu2020semi}: TCF aims to enhance the representation of a user or item by combining its textual representation with features extracted from the interaction data.

   \item  \textbf{Text-enhanced Domain Adaptation Recommendation (TDAR) }~\cite{yu2020semi}: TDAR is a semi-supervised domain adaptation-based cross-domain recommendation model that considers TCF as a base model to align embedding of entities (users or items) across domain based on textual features. 

   \item \AS{\textbf{Cross-Domain Recommendation via Aspect Transfer Network (CATN \& CATN-separate)} ~\cite{zhao2020catn}: This approach aims to extract user and item aspects from the review data, and learn aspect correlations across domains through a global aspect representation with attention (CATN-separate). Further, the user's aspect representation is enhanced by incorporating auxiliary reviews from like-minded users (CATN).}
\end{itemize}

\subsection{Evaluation Metrics}
%We evaluate the performance of the models using $F_1$-score and Normalized Discounted Cumulative Gain (NDCG) measures. $F_1$-score is the harmonic mean of precision and recall where precision measures the fraction of relevant items among the recommended items while recall measures the fraction of relevant items that are recommended among the total relevant items. Thus this is a measure of relevance. NDCG is a measure of ranking quality. It considers the position of item in the ranking for evaluating the model performance. We evaluate the model by recommending Top-k items where $k \in \{2, 5, 10, 15, 20\}$ for each user and use the average metric value of all users for final evaluation of model.
\adamya{To assess the models' performance, we utilized two metrics: the $F_1$-score and Normalized Discounted Cumulative Gain (NDCG). The $F_1$-score represents the harmonic mean of precision and recall. Precision measures the proportion of relevant items among the recommended items, while recall indicates the fraction of relevant items recommended among the total relevant items. Conversely, NDCG is a metric evaluating the ranking quality, considering the item's position in the recommendation set to evaluate the model's performance.} % We evaluated the models' performance using two evaluation measures: $F_1$-score and Normalized Discounted Cumulative Gain (NDCG). The $F_1$-score is the harmonic mean of precision and recall, where precision denotes the fraction of relevant items among the recommended items and recall is the fraction of relevant items recommended among the total relevant items. On the other hand, NDCG is a measure of ranking quality that considers the position of the item in the recommendation set for evaluating the model performance. %We evaluated the model performance in a top-k recommendation scenario for different values of $k \in \{2, 5, 10, 15, 20\}$. 
The model performance is evaluated in a top-\textit{k} recommendation scenario, where we considered different values of $k$ from the set $\{2, 5, 10, 15, 20\}$.

\subsection{Experimental Setting}
\label{sec:Experimental-setting}
\AS{For a fair comparison of our proposed approach with the existing benchmark techniques, we performed hyperparameter tuning of all models over the considered datasets. For each algorithm, we have followed the tuning approach suggested by Yu et al.~\cite{yu2020semi}. Firstly, we employ a random train-test-validation split of the datasets in the ratio of 8:1:1. We train each of the comparing algorithms over the train set, and fine-tune their corresponding parameters using the validation set. We evaluate each of the inducted models over the test set.} 

\begin{table}[ht]
    \centering
    \renewcommand{\arraystretch}{1.2}
    \caption{Optimal parameters for single-domain recommendation models.}
    \adjustbox{max width=0.5\linewidth}{
    \begin{tabular}{llll}
    \hline
    Dataset & Model & $\eta$ & $\lambda$ \\
    \hline
    \multirow{4}{*}{Cell Phones \& Accessories} 
        & TMN   & 0.0003 & 0.3       \\
        & TCF   & 0.1    & 0.007     \\
        & VCF   & 0.1    & 0.1       \\
        & FCF   & 0.09   & 0.01      \\
        \hline    
    \multirow{4}{*}{Electronics} 
        & TMN   & 0.0001 & 0.3       \\
        & TCF   & 0.05   & 0.005     \\
        & VCF   & 0.2    & 0.0009    \\
        & FCF   & 0.06   & 0.009     \\
        \hline
    \multirow{4}{*}{Movies \& TV} 
        & TMN   & 0.0001 & 0.004     \\
        & TCF   & 0.01   & 0.0008    \\
        & VCF   & 0.8    & 0.001     \\
        & FCF   & 0.05   & 0.004     \\
        \hline
    \multirow{4}{*}{Video Games} 
        & TMN   & 0.0001 & 0.08      \\
        & TCF   & 0.02   & 0.0009    \\
        & VCF   & 1      & 0.4       \\
        & FCF   & 0.1    & 0.002     \\
        \hline
    \end{tabular}
    }
    \label{tab:optimalHyperSDR}
\end{table}

\begin{figure*}
    \centering
        \begin{subfigure}[b]{0.49\textwidth}
            \centering
            \includegraphics[width=\textwidth, height= 6.5cm]{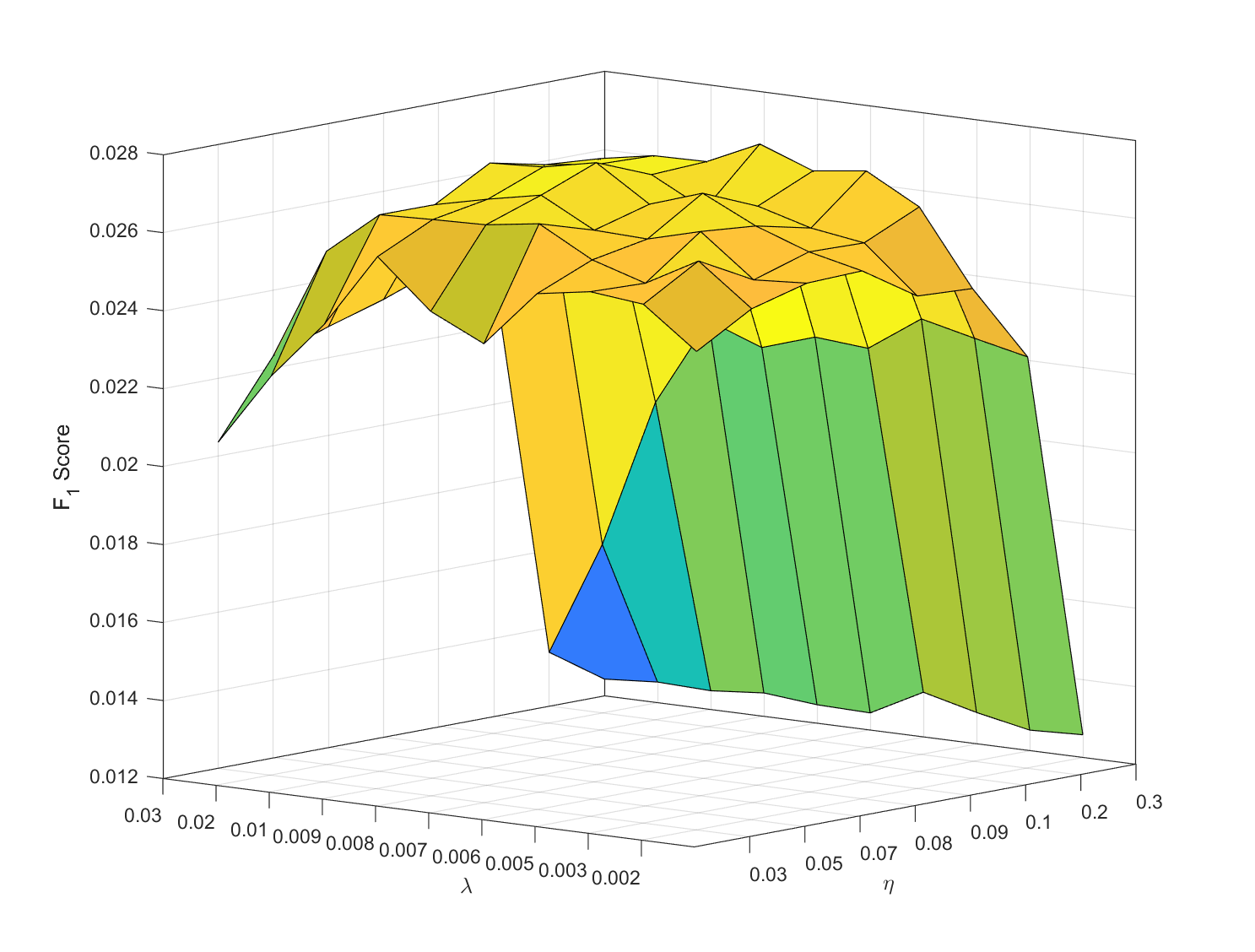}
            \caption{TCF}
            \label{fig:TCF_C}
        \end{subfigure}
        \hfill
        \begin{subfigure}[b]{0.49\textwidth}
            \centering
            \includegraphics[width=\textwidth, height= 6.5cm]{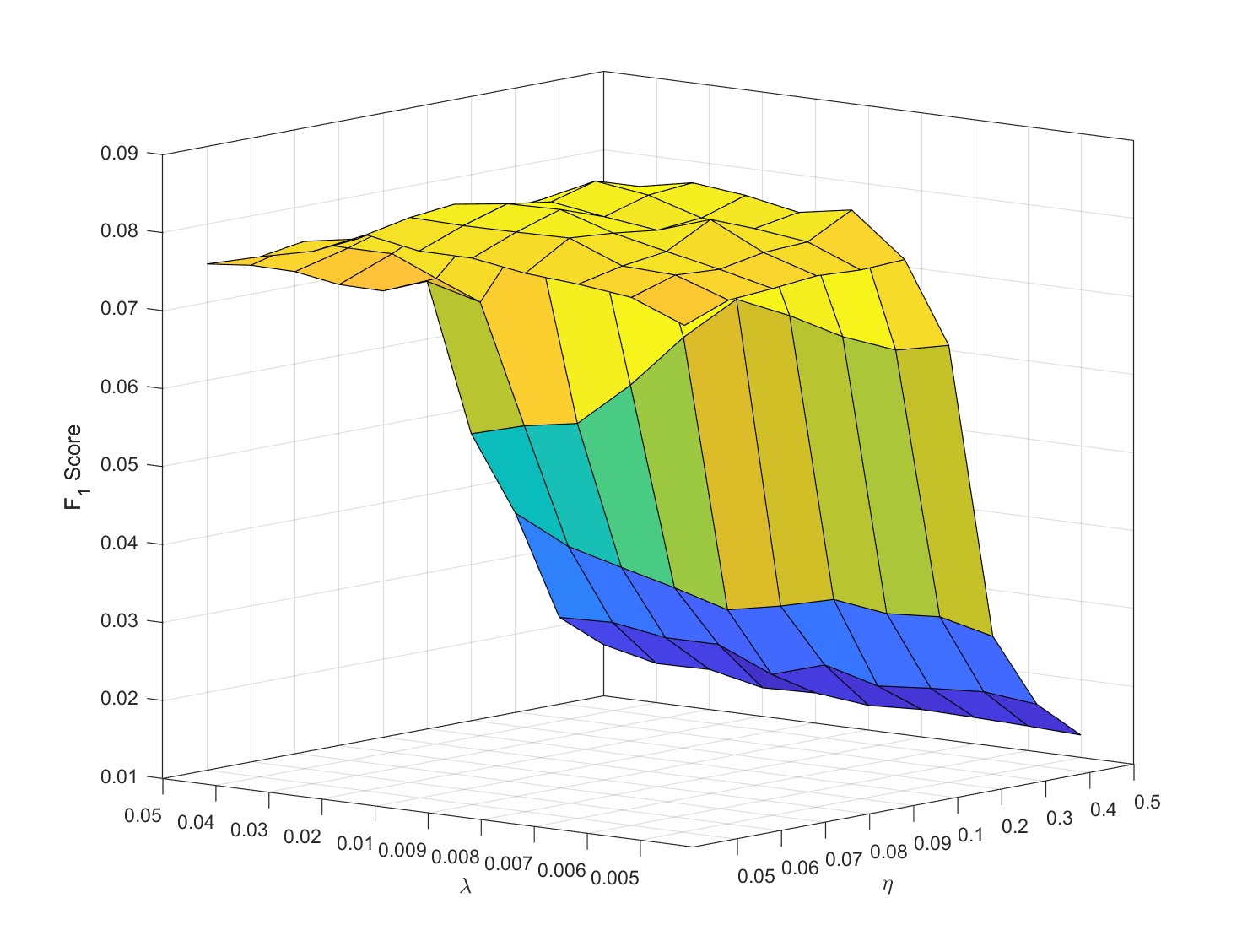}
            \caption{FCF}
            \label{fig:FCF_C}
        \end{subfigure}
   \caption{Effect of learning rate ($\eta$) and regularization coefficient ($\lambda$) on TCF and FCF over Cell Phones and Accessories dataset.}
    \label{fig:HyperTuneSDR}
\end{figure*}

\AS{We have employed a coarse and fine-grain tuning approach to obtain the optimal values of learning rate $\eta$ and regularization coefficient $\lambda$ for single-domain recommendation models. Initially, we tune $\eta$ and $\lambda$ in a broad range of \{0.0001, 0.001, 0.01, 0.1, 1\} $\otimes$ \{0.0001, 0.001, 0.01, 0.1, 1\}, where $\otimes$ represents Cartesian product. Further, for example, if a specific model performs best for $\eta = 0.001$ and $\lambda = 0.01$, then we fine-tune it in $\{0.0007, 0.0009, 0.001, 0.003, 0.005, 0.007\} \otimes \{0.007, 0.009, 0.01, 0.03,0.05\}$. We have applied the same approach to the CATN model. The optimal values of parameters for single domain recommendation models are provided in Table~\ref{tab:optimalHyperSDR}. Figure \ref{fig:HyperTuneSDR} shows the impact of varying the regularization coefficient ($\lambda$) and learning rate ($\eta$) on the performance of the TCF and FCF models using the Cell Phones and Accessories dataset.}

\begin{table}[!ht]
    \centering
    \renewcommand{\arraystretch}{1.2}
    \caption{Optimal parameters for cross-domain recommendation models.}
    \adjustbox{max width=0.6\linewidth}{
    \begin{tabular}{llllll}
    \hline
    Scenario & Model & $\eta^+$ & $\eta^-$ & $\eta$ & $\lambda$ \\
    \hline
    \multirow{5}{*}{EC} 
        & TDAR          & 0.01   & 0.0001 & 0.9    & 0.002   \\
        & CATN          & -      & -      & 0.1    & 0.0001  \\
        & CATN-separate & -      & -      & 0.001  & 1       \\
        & VDAR          & 0.01   & 0.6    & 0.001  & 0.0007  \\
        & FDAR          & 0.01   & 0.005  & 0.005  & 0.0009  \\
        \hline
    \multirow{5}{*}{EV} 
        & TDAR          & 0.0001 & 0.0001 & 0.09   & 0.02    \\
        & CATN          & -      & -      & 0.9    & 0.3     \\
        & CATN-separate & -      & -      & 0.08   & 0.9     \\
        & VDAR          & 0.1    & 0.1    & 0.0008 & 0.003   \\
        & FDAR          & 0.1    & 0.0001 & 0.009  & 0.0002  \\
        \hline
    \multirow{5}{*}{MV} 
        & TDAR          & 0.0001 & 0.0001 & 1      & 0.003   \\
        & CATN          & -      & -      & 0.6    & 0.09    \\
        & CATN-separate & -      & -      & 0.2    & 0.0001  \\
        & VDAR          & 0.0001 & 0.1    & 0.001  & 0.3     \\
        & FDAR          & 0.1    & 0.0001 & 0.009  & 0.0002  \\
        \hline
    \end{tabular}
    \label{tab:optimalHyperCDR}
    }
\end{table}

\AS{For cross-domain recommendation models, we have tuned the values of $\eta^+, \eta^-, \eta^t, \text{and } \lambda$. Firstly, to obtain the optimal value of $\eta^+$, we set $\eta^s, \eta^t, \text{and } \eta^-$ to $0$. Similarly, for $\eta^-$, we set $\eta^+$ as the optimal value obtained from the previous step and the rest as $0$. We have determined the best values for these parameters in the same manner as discussed above with a coarse grain range of ${0.0001,0.001,0.01,0.1,1}$. Further, we set $\eta^s$ and $\lambda^s$ as the optimal value obtained for source domain through the corresponding single domain recommendations models.  
Lastly, for the target domain, we tune $\eta^t$ and $\lambda^t$ using the approach defined for single-domain models. The combination of hyperparameters corresponding to the best average $F_{1}$-scores of top-k recommendations ($k$ = $\{2, 5, 10, 15, 20\}$) is considered as an optimal set. The optimal set of parameters for cross-domain recommendation models is provided in  Table \ref{tab:optimalHyperCDR}. We have reported the results of each algorithm as the average of three runs over the test set.}

\subsection{Empirical Analysis}
\AS{
In this section, we conduct a comparative analysis of the proposed models against the state-of-the-art methods using the benchmark datasets. \VB{For each evaluation metric, ``$\uparrow$" indicates that a higher value is better. The best performance among the comparing algorithms is highlighted in boldface.} We assessed the performance of these models in both single-domain and cross-domain recommendation scenarios. The textual and visual features are utilized as side information in the single-domain recommendation scenario to enhance the representation of entities. In the cross-domain recommendation scenario, these features are used for domain adaptation.  The results of our experimental analysis for the single-domain recommendation scenario, evaluated using the $F_1$ and $NDCG$ measures, are presented in Table \ref{tab:f1SDR} and \ref{tab:ndcgSDR}, respectively.  The results are presented for various top-k values. }

\begin{table}[h]
    \renewcommand{\arraystretch}{1.2}
    \centering	
    % \captionsetup{font=scriptsize,justification=centering}
    \caption{Experimental results of each comparing algorithms (mean rank) for single-domain recommendations in terms of $F_1@k$ score.}
    \adjustbox{max width=\linewidth}{
    \begin{tabular}{@{}clllllllllll@{}}
    \toprule
    \textbf{Dataset} & \textbf{Model} &  \textbf{$F_1$@2 $(\uparrow)$} & \textbf{} & \textbf{$F_1$@5 $(\uparrow)$}  & \textbf{} & \textbf{$F_1$@10 $(\uparrow)$}   & \textbf{} & \textbf{$F_1$@15 $(\uparrow)$} & \textbf{} & \textbf{$F_1$@20 $(\uparrow)$}   & \textbf{} \\ \midrule
    
    \multirow{4}{*}{\textbf{Cell Phones \& Accessories}} 
    & \textbf{TMN} \cite{yu2020semi} & 0.01659 & 3 & 0.01245 & 3 & 0.00926 & 3 & 0.00767 & 3 & 0.00668 & 3 \\
    & \textbf{TCF} \cite{yu2020semi} & 0.04163 & 2 & 0.03054 & 2 & 0.02249 & 2 & 0.01835 & 2 & 0.01565 & 2 \\
    & \textbf{VCF} & 0.00165 & 4 & 0.00136 & 4 & 0.00103 & 4 & 0.00079 & 4 & 0.00083 & 4 \\
    & \textbf{FCF} & \textbf{0.14052} & 1 & \textbf{0.10332} & 1 & \textbf{0.07219} & 1 & \textbf{0.05657} & 1 & \textbf{0.04751} & 1 \\ \midrule

    \multirow{4}{*}{\textbf{Electronics}} 
    & \textbf{TMN} \cite{yu2020semi} & 0.00918 & 3 & 0.00734 & 3 & 0.00642 & 3 & 0.00576 & 3 & 0.00530 & 3 \\
    & \textbf{TCF} \cite{yu2020semi} & 0.03383 & 2 & 0.02782 & 2 & 0.02142 & 2 & 0.01796 & 2 & 0.01565 & 2 \\
    & \textbf{VCF} & 0.00219 & 4 & 0.00186 & 4 & 0.00182 & 4 & 0.00163 & 4 & 0.00157 & 4 \\
    & \textbf{FCF} & \textbf{0.03578} & 1 & \textbf{0.03344} & 1 & \textbf{0.02837} & 1 & \textbf{0.02448} & 1 & \textbf{0.02180} & 1 \\ \midrule
    
    \multirow{4}{*}{\textbf{Movies \& TV}} 
    & \textbf{TMN} \cite{yu2020semi} & 0.01642 & 3 & 0.01597 & 3 & 0.01412 & 3 & 0.01320 & 3 & 0.01252 & 3 \\
    & \textbf{TCF} \cite{yu2020semi} & 0.06104 & 2 & 0.06018 & 2 & 0.05489 & 2 & 0.05012 & 2 & 0.04594 & 2 \\
    & \textbf{VCF} & 0.00610 & 4 & 0.00491 & 4 & 0.00367 & 4 & 0.00358 & 4 & 0.00342 & 4 \\
    & \textbf{FCF} & \textbf{0.15340} & 1 & \textbf{0.12840} & 1 & \textbf{0.09631} & 1 & \textbf{0.07825} & 1 & \textbf{0.06665} & 1 \\ \midrule
    
    \multirow{4}{*}{\textbf{Video Games}} 
    & \textbf{TMN} \cite{yu2020semi} & 0.01687 & 4 & 0.01426 & 4 & 0.01412 & 4 & 0.01293 & 4 & 0.01251 & 3 \\
    & \textbf{TCF} \cite{yu2020semi} & 0.06969 & 2 & 0.06503 & 2 & 0.05298 & 2 & 0.04550 & 2 & 0.04094 & 2 \\
    & \textbf{VCF} & 0.01752 & 3 & 0.01622 & 3 & 0.01581 & 3 & 0.01345 & 3 & 0.01076 & 4 \\
    & \textbf{FCF} & \textbf{0.21693} & 1 & \textbf{0.15970} & 1 & \textbf{0.10981} & 1 & \textbf{0.08782} & 1 & \textbf{0.07383} & 1 \\ 
    \bottomrule
    
    \end{tabular}
    }
    \label{tab:f1SDR}
\end{table}

\begin{table}[!h]
    \renewcommand{\arraystretch}{1.2}
    \centering	
    % \captionsetup{font=scriptsize,justification=centering}
    \caption{Experimental results of each comparing algorithms (mean rank) for single-domain recommendations in terms of $NDCG@k$ score.}
    \adjustbox{max width=\linewidth}{
    \begin{tabular}{@{}clllllllllll@{}}
    \toprule
    \textbf{Dataset} & \textbf{Model} & \textbf{$NDCG$@2 $(\uparrow)$} & \textbf{} & \textbf{$NDCG$@5 $(\uparrow)$}  & \textbf{} & \textbf{$NDCG$@10 $(\uparrow)$}   & \textbf{} & \textbf{$NDCG$@15 $(\uparrow)$} & \textbf{} & \textbf{$NDCG$@20 $(\uparrow)$}   & \textbf{} \\ \midrule
    
    \multirow{4}{*}{\textbf{Cell Phones \& Accessories}}
    & \textbf{TMN} \cite{yu2020semi} & 0.02116 & 3 & 0.02440 & 3 & 0.02763 & 3 & 0.02974 & 3 & 0.03134 & 3 \\
    & \textbf{TCF} \cite{yu2020semi} & 0.05517 & 2 & 0.06265 & 2 & 0.07006 & 2 & 0.07444 & 2 & 0.07752 & 2 \\
    & \textbf{VCF} & 0.00191 & 4 & 0.00233 & 4 & 0.00271 & 4 & 0.00283 & 4 & 0.00325 & 4 \\
    & \textbf{FCF} & \textbf{0.18403} & 1 & \textbf{0.21507} & 1 & \textbf{0.23585} & 1 & \textbf{0.24673} & 1 & \textbf{0.25511} & 1 \\ \midrule
    
    \multirow{4}{*}{\textbf{Electronics}}
    & \textbf{TMN} \cite{yu2020semi} & 0.01248 & 3 & 0.01291 & 3 & 0.01531 & 3 & 0.01702 & 3 & 0.01835 & 3 \\
    & \textbf{TCF} \cite{yu2020semi} & 0.04368 & 2 & 0.04844 & 2 & 0.05415 & 2 & 0.05795 & 2 & 0.06062 & 2 \\
    & \textbf{VCF} & 0.00266 & 4 & 0.00303 & 4 & 0.00396 & 4 & 0.00450 & 4 & 0.00502 & 4 \\
    & \textbf{FCF} & \textbf{0.04534} & 1 & \textbf{0.05446} & 1 & \textbf{0.06493} & 1 & \textbf{0.07112} & 1 & \textbf{0.07573} & 1 \\ \midrule
    
    \multirow{4}{*}{\textbf{Movies \& TV}}
    & \textbf{TMN} \cite{yu2020semi} & 0.02104 & 3 & 0.02609 & 3 & 0.03169 & 3 & 0.03615 & 3 & 0.03986 & 3 \\
    & \textbf{TCF} \cite{yu2020semi} & 0.07963 & 2 & 0.09802 & 2 & 0.12065 & 2 & 0.13564 & 2 & 0.14655 & 2 \\
    & \textbf{VCF} & 0.00757 & 4 & 0.00818 & 4 & 0.00898 & 4 & 0.01031 & 4 & 0.01141 & 4 \\
    & \textbf{FCF} & \textbf{0.19805} & 1 & \textbf{0.22527} & 1 & \textbf{0.24938} & 1 & \textbf{0.26264} & 1 & \textbf{0.27172} & 1 \\ \midrule
    
    \multirow{4}{*}{\textbf{Video Games}}
    & \textbf{TMN} \cite{yu2020semi} & 0.01977 & 4 & 0.02535 & 4 & 0.03408 & 4 & 0.03932 & 4 & 0.04412 & 3 \\
    & \textbf{TCF} \cite{yu2020semi} & 0.08709 & 2 & 0.11597 & 2 & 0.13855 & 2 & 0.15250 & 2 & 0.16410 & 2 \\
    & \textbf{VCF} & 0.02062 & 3 & 0.02765 & 3 & 0.03678 & 3 & 0.04092 & 3 & 0.04173 & 4 \\
    & \textbf{FCF} & \textbf{0.27704} & 1 & \textbf{0.32252} & 1 & \textbf{0.35078} & 1 & \textbf{0.36946} & 1 & \textbf{0.38221} & 1 \\
    \bottomrule
    
    \end{tabular}
    }
    \label{tab:ndcgSDR}
\end{table}

\AS{
It is evident from the tables that our proposed FCF model, which combines visual, textual, and latent features derived from implicit feedback, significantly enhances recommendation accuracy compared to models that rely solely on textual features or combinations of latent features with either textual or visual features. Textual and visual features alone are often insufficient in capturing complex interaction patterns between users and items. Visual features primarily capture surface-level, aesthetic, or contextual characteristics, such as design, color, and form factor, but they do not fully convey the deeper, semantic meaning of an item or its relevance to a user's specific needs. Similarly, textual features, though rich in semantic information, can lack contextual or aesthetic insights. While they provide detailed descriptions, reviews, or specifications, they may not convey visual aspects of an item, such as its design or appeal.  Therefore, the synergistic combination of visual features with the semantic depth of textual features and the latent features uncovering the patterns in implicit feedback enables the FCF model to construct a more comprehensive and robust entity representation. This holistic representation allows the model to better understand user preferences and item characteristics, resulting in superior performance in terms of recommendation accuracy.}

\begin{table}[ht!]
    \renewcommand{\arraystretch}{1.1}
    \centering	
    % \captionsetup{font=scriptsize,justification=centering}
    \caption{Experimental results of each comparing algorithms (mean rank) for cross-domain recommendations in terms of $F_1@k$ score.}
    \adjustbox{max width=\linewidth}{
    \begin{tabular}{@{}clllllllllll@{}}
    \toprule
    \textbf{Scenario} & \textbf{Model} & \textbf{$F_1$@2 $(\uparrow)$} & \textbf{} & \textbf{$F_1$@5 $(\uparrow)$}  & \textbf{} & \textbf{$F_1$@10 $(\uparrow)$} & \textbf{} & \textbf{$F_1$@15 $(\uparrow)$} & \textbf{} & \textbf{$F_1$@20 $(\uparrow)$} & \textbf{} \\ \midrule

    \multirow{5}{*}{\textbf{EC}} 
    & \textbf{TDAR} \cite{yu2020semi} & 0.04103 & 4 & 0.03171 & 4 & 0.02424 & 3 & 0.02046 & 3 & 0.01797 & 3 \\
    & \textbf{CATN} \cite{zhao2020catn} & 0.08286 & 2 & 0.05559 & 2 & 0.04037 & 2 & 0.03429 & 2 & 0.03102 & 2 \\
    & \textbf{CATN-separate} \cite{zhao2020catn} & 0.07403 & 3 & 0.04217 & 3 & 0.02412 & 4 & 0.01692 & 4 & 0.01304 & 4 \\
    & \textbf{VDAR}          & 0.00127 & 5 & 0.00152 & 5 & 0.00146 & 5 & 0.00123 & 5 & 0.00109 & 5 \\
    & \textbf{FDAR} & \textbf{0.14884} & 1 & \textbf{0.10582} & 1 & \textbf{0.07370} & 1 & \textbf{0.05758} & 1 & \textbf{0.04845} & 1 \\ \midrule
    
    \multirow{5}{*}{\textbf{EV}}
    & \textbf{TDAR} \cite{yu2020semi} & 0.07306 & 4 & 0.06715 & 3 & 0.05616 & 2 & 0.04873 & 2 & 0.04333 & 2 \\
    & \textbf{CATN} \cite{zhao2020catn} & 0.14477 & 2 & 0.08228 & 2 & 0.04714 & 3 & 0.03307 & 3 & 0.02548 & 3 \\
    & \textbf{CATN-separate} \cite{zhao2020catn} & 0.12416 & 3 & 0.06514 & 4 & 0.04506 & 4 & 0.01651 & 4 & 0.01544 & 4 \\
    & \textbf{VDAR}          & 0.01787 & 5 & 0.01581 & 5 & 0.01512 & 5 & 0.01349 & 5 & 0.01085 & 5 \\
    & \textbf{FDAR} & \textbf{0.22126} & 1 & \textbf{0.15929} & 1 & \textbf{0.11030} & 1 & \textbf{0.08963} & 1 & \textbf{0.07578} & 1 \\ \midrule
    
    \multirow{5}{*}{\textbf{MV}}
    & \textbf{TDAR} \cite{yu2020semi} & 0.07235 & 4 & 0.06695 & 2 & 0.05725 & 2 & 0.04919 & 2 & 0.04383 & 2 \\
    & \textbf{CATN} \cite{zhao2020catn} & 0.08472 & 2 & 0.04815 & 3 & 0.02758 & 3 & 0.01935 & 3 & 0.01491 & 3 \\
    & \textbf{CATN-separate} \cite{zhao2020catn} & 0.08247 & 3 & 0.04320 & 4 & 0.02450 & 4 & 0.01910 & 4 & 0.01487 & 4 \\
    & \textbf{VDAR}          & 0.01833 & 5 & 0.01580 & 5 & 0.01504 & 5 & 0.01387 & 5 & 0.01134 & 5 \\
    & \textbf{FDAR} & \textbf{0.22106} & 1 & \textbf{0.15941} & 1 & \textbf{0.11038} & 1 & \textbf{0.08964} & 1 & \textbf{0.07583} & 1 \\ 
    \bottomrule
    
    \end{tabular}
    }
    \label{tab:f1CDR}
\end{table}

\begin{table}[ht!]
    \renewcommand{\arraystretch}{1.1}
    \centering	
    % \captionsetup{font=scriptsize,justification=centering}
    \caption{Experimental results of each comparing algorithms (mean rank) for cross-domain recommendations in terms of $NDCG@k$ score.}
    \adjustbox{max width=\linewidth}{
    \begin{tabular}{@{}clllllllllll@{}}
    \toprule
    \textbf{Scenario} & \textbf{Model} & \textbf{$NDCG$@2 $(\uparrow)$} & \textbf{} & \textbf{$NDCG$@5 $(\uparrow)$}  & \textbf{} & \textbf{$NDCG$@10 $(\uparrow)$} & \textbf{} & \textbf{$NDCG$@15 $(\uparrow)$} & \textbf{} & \textbf{$NDCG$@20 $(\uparrow)$} & \textbf{} \\ \midrule

    \multirow{5}{*}{\textbf{EC}}
    & \textbf{TDAR} \cite{yu2020semi} & 0.05388 & 4 & 0.06289 & 4 & 0.07214 & 4 & 0.07796 & 4 & 0.08263 & 4 \\
    & \textbf{CATN} \cite{zhao2020catn} & 0.08625 & 2 & 0.08806 & 2 & 0.08915 & 2 & 0.08989 & 2 & 0.09448 & 2 \\
    & \textbf{CATN-separate} \cite{zhao2020catn} & 0.06113 & 3 & 0.07109 & 3 & 0.07607 & 3 & 0.07836 & 3 & 0.08429 & 3 \\
    & \textbf{VDAR}          & 0.00143 & 5 & 0.00223 & 5 & 0.00309 & 5 & 0.00344 & 5 & 0.00373 & 5 \\
    & \textbf{FDAR} & \textbf{0.19377} & 1 & \textbf{0.22183} & 1 & \textbf{0.24269} & 1 & \textbf{0.25343} & 1 & \textbf{0.26213} & 1 \\ \midrule
    
    \multirow{5}{*}{\textbf{EV}}
    & \textbf{TDAR} \cite{yu2020semi} & 0.09091 & 4 & 0.11996 & 4 & 0.14605 & 4 & 0.16162 & 4 & 0.17281 & 4 \\
    & \textbf{CATN} \cite{zhao2020catn} & 0.19343 & 2 & 0.20122 & 2 & 0.20925 & 2 & 0.21042 & 2 & 0.21885 & 2 \\
    & \textbf{CATN-separate} \cite{zhao2020catn} & 0.17832 & 3 & 0.18124 & 3 & 0.18875 & 3 & 0.19316 & 3 & 0.19879 & 3 \\
    & \textbf{VDAR}          & 0.02073 & 5 & 0.02710 & 5 & 0.03594 & 5 & 0.04081 & 5 & 0.04166 & 5 \\
    & \textbf{FDAR} & \textbf{0.28323} & 1 & \textbf{0.32545} & 1 & \textbf{0.35433} & 1 & \textbf{0.37591} & 1 & \textbf{0.38927} & 1 \\ \midrule
    
    \multirow{5}{*}{\textbf{MV}}
    & \textbf{TDAR} \cite{yu2020semi} & 0.09058 & 4 & 0.12019 & 3 & 0.14838 & 2 & 0.16322 & 2 & 0.17517 & 2 \\
    & \textbf{CATN} \cite{zhao2020catn} & 0.12366 & 2 & 0.13786 & 2 & 0.13973 & 3 & 0.14003 & 3 & 0.17202 & 3 \\
    & \textbf{CATN-separate} \cite{zhao2020catn} & 0.10707 & 3 & 0.11893 & 4 & 0.12923 & 4 & 0.13122 & 4 & 0.13927 & 4 \\
    & \textbf{VDAR}          & 0.02129 & 5 & 0.02735 & 5 & 0.03607 & 5 & 0.04164 & 5 & 0.04289 & 5 \\
    & \textbf{FDAR} & \textbf{0.28307} & 1 & \textbf{0.32557} & 1 & \textbf{0.35446} & 1 & \textbf{0.37598} & 1 & \textbf{0.38941} & 1 \\
    \bottomrule
        
    \end{tabular}
    }
    \label{tab:ndcgCDR}
\end{table}

\AS{To evaluate the performance of our proposed cross-domain recommendation approaches FDAR,  we have considered three configurations:  \textit{Electronics} $\rightarrow$ \textit{Cell Phones and Accessories} (EC), \textit{Electronics} $\rightarrow$ \textit{Video Games} (EV), and \textit{Movies and TV} $\rightarrow$ \textit{Video Games} (MV).  Here, $A \rightarrow B$ indicates that dataset $A$ serves as the source domain, while dataset $B$ is used as the target domain. Table~\ref{tab:f1CDR} and \ref{tab:ndcgCDR} report the results related to this experiment. Similar to the single-domain recommendation scenario, we observe similar behavior across models that use visual or textual features alone or in combination. The proposed approach FDAR, which combines textual, visual, and latent features for the domain alignment, outperforms other methods. This superior performance can be attributed to the synergy between visual and textual representations. For example, in the \textit{Electronics} domain, the visual features of a smartphone can help highlight its compatibility with accessories. Meanwhile, textual descriptions of smartphones provide additional details, such as the model number, screen size, and technical specifications, while accessory descriptions specify compatibility and features. By combining both visual and textual information, FDAR creates a more comprehensive representation of the items, improving the alignment of user-item interactions across domains and resulting in more accurate cross-domain recommendations.}

\begin{table}[]
    \renewcommand{\arraystretch}{1.2}
    \centering	
    \caption{Summary of the Friedman statistics ($F_F$) and critical values for the $F1@5$ and $NDCG@5$ evaluation metrics. $\mathcal{N}$: Number of datasets/scenarios, $\mathcal{K}$: Number of comparing algorithms.}
    \adjustbox{max width=0.75\linewidth}{
    \begin{tabular}{lllclccl}
        \toprule
        & Metric   & $F_F$ & Critical Value ($\alpha = 0.05$) & $q_\alpha$ & \multicolumn{1}{l}{$\mathcal{N}$} & \multicolumn{1}{l}{$\mathcal{K}$} & CD \\
        \midrule
        & $F_1$@5  & 37 & & & &\\
        \multirow{-2}{*}{SDR} 
        & $NDCG$@5 & 37 & \multirow{-2}{*}{3.8625} & \multirow{-2}{*}{2.569} & \multirow{-2}{*}{4}   & \multirow{-2}{*}{4} & \multirow{-2}{*}{2.3452} \\
        \midrule
        & $F_1$@5  & 16    &                                 &                            &                       &\\
        \multirow{-2}{*}{CDR}
        & $NDCG$@5 & 43  & \multirow{-2}{*}{3.8379}  & \multirow{-2}{*}{2.728} & \multirow{-2}{*}{3}   & \multirow{-2}{*}{5} & \multirow{-2}{*}{3.5218} \\
        \bottomrule
    \end{tabular}
    }
\label{tab:FriedNem}
\end{table}

\AS{To further validate our experiment results, we conducted a statistical analysis to compare the performance of the algorithms for both single-domain recommendations (SDR) and cross-domain recommendations (CDR). We applied the \textit{Friedman test}, which is a suitable statistical test for comparing more than two algorithms across multiple datasets~\cite{demvsar2006statistical}. 
%To estimate the Friedman statistic, $F_F$, we first calculate the $\chi^2_F$ as
%\begin{equation}
%    \chi^2_F = \frac{12\mathcal{N}}{\mathcal{K}(\mathcal{K}+1)} \sum_{j} R_j^2 - \frac{\mathcal{K}(\mathcal{K}+1)^2}{4},
%\end{equation}
%\AS{\noindent where $R_j$ is the average rank of $j_th$ algorithm over the $\mathcal{N}$ datasets in terms of the concerned evaluation metric, and$\mathcal{K}$ is the number of comparing algorithms.}
%\noindent \AS{Further, $F_F$ is computed as,}
%\begin{equation}
%    F_F = \frac{(\mathcal{N}-1)\chi^2_F}{\mathcal{N}(\mathcal{K}-1)-\chi^2_F}.
%\end{equation}
The summary of the Friedman statistics analysis is presented in Table \ref{tab:FriedNem}. For both SDR and CDR, at the significance level of $\alpha = 0.05$, the \textit{Friedman test} rejects the null hypothesis that there is no significant difference in the performance of the comparing algorithms over the considered evaluation metrics. To further assess the performance of our proposed approach against the comparing algorithms, we also apply the \textit{Nemenyi post-hoc test}~\cite{demvsar2006statistical}.} 

\begin{figure}[h]
    \centering
        \begin{subfigure}[b]{0.49\textwidth}
            \centering
            \includegraphics[width=\textwidth, trim=60 68 60 72, clip]{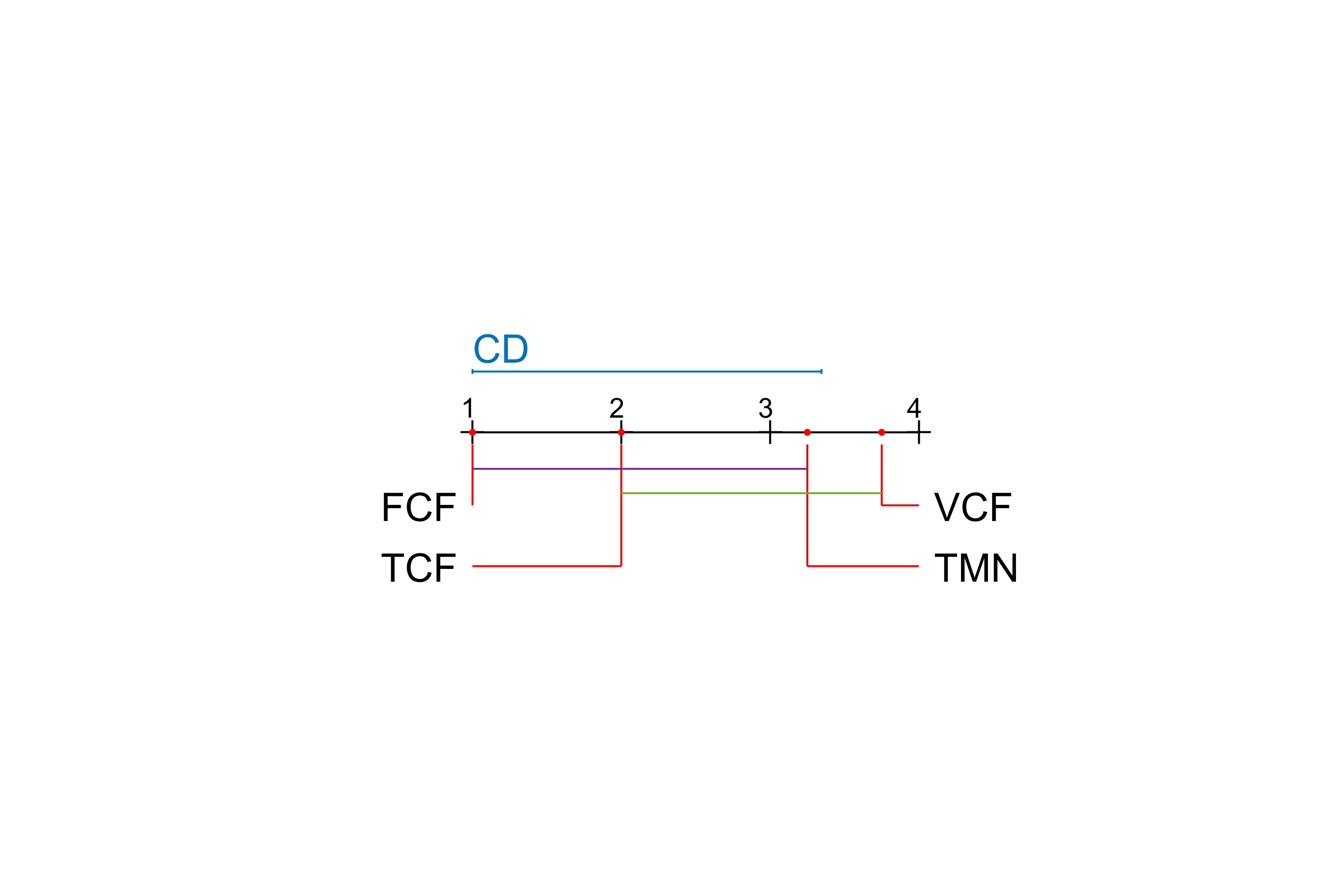}
            \caption{$F_1@5$}
            \label{fig:f1SDR}
        \end{subfigure}
        \hfill
        \begin{subfigure}[b]{0.49\textwidth}
            \centering
            \includegraphics[width=\textwidth, trim=60 68 60 72, clip]{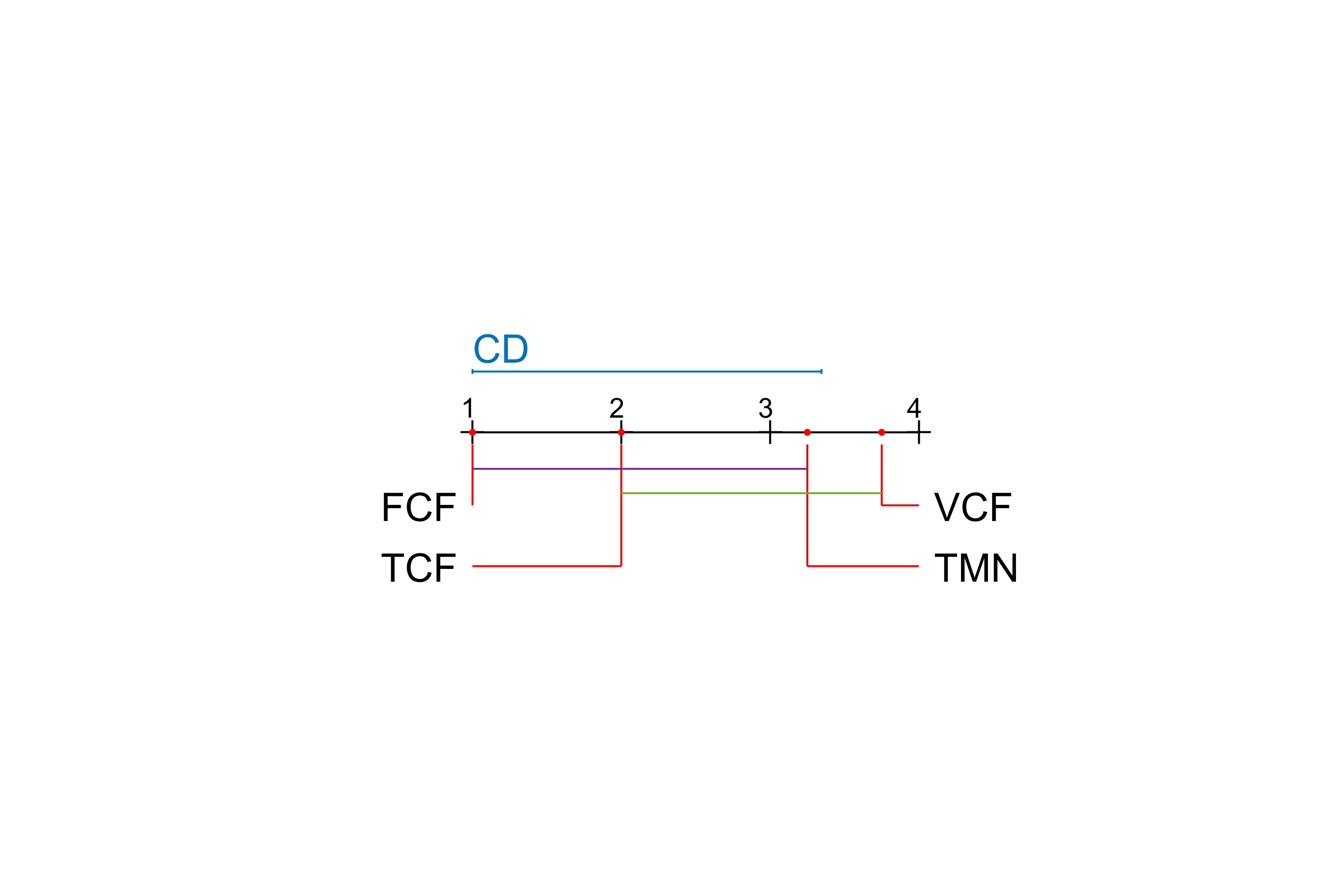}
            \caption{$NDCG@5$}
            \label{fig:ndcgSDR}
        \end{subfigure}
        \begin{subfigure}[b]{0.49\textwidth}
            \centering
            \includegraphics[width=\textwidth, trim=45 58 25 68, clip]{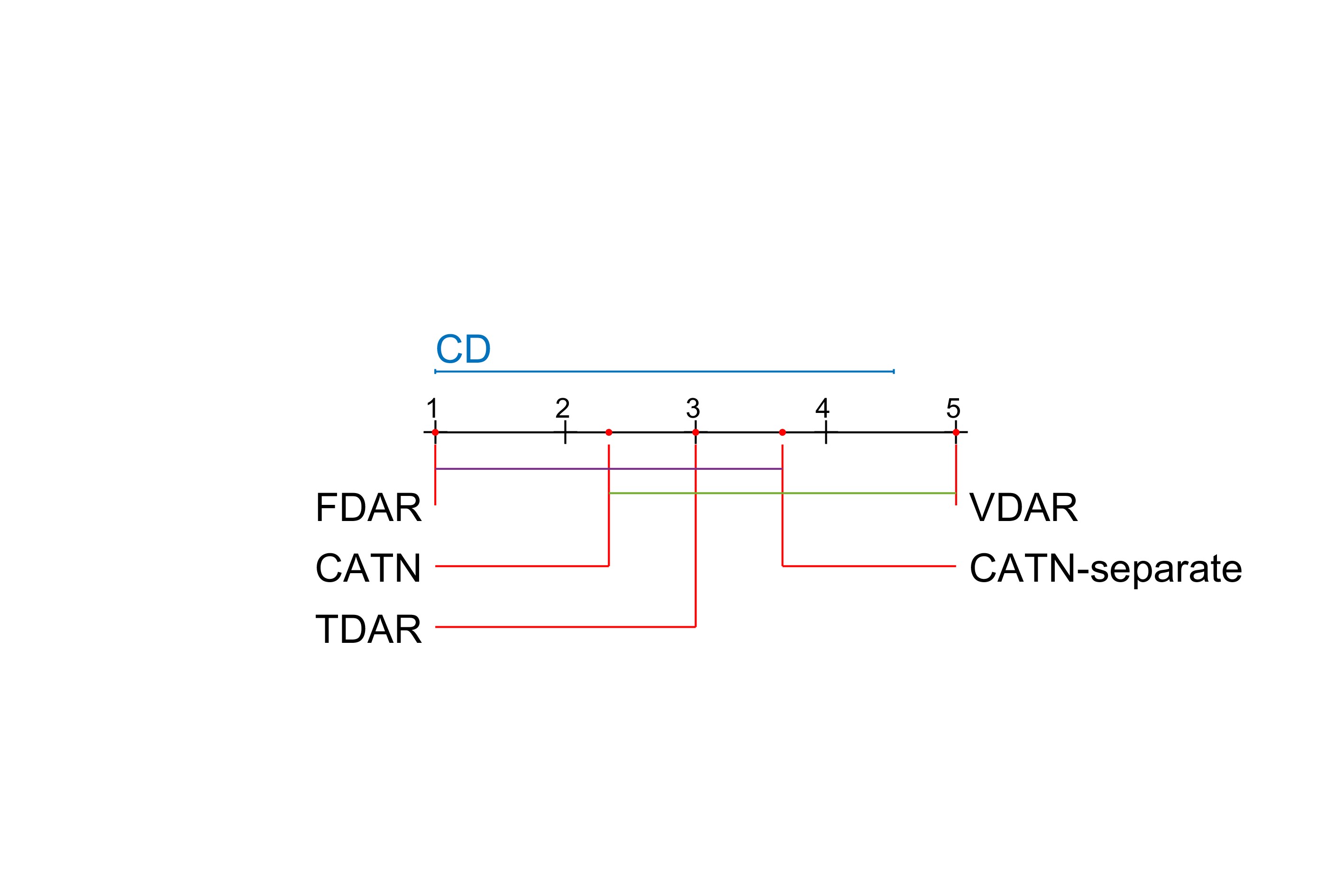}
            \caption{$F_1@5$}
            \label{fig:f1CDR}
        \end{subfigure}
        \hfill
        \begin{subfigure}[b]{0.49\textwidth}
            \centering
            \includegraphics[width=\textwidth, trim=30 58 40 68, clip]{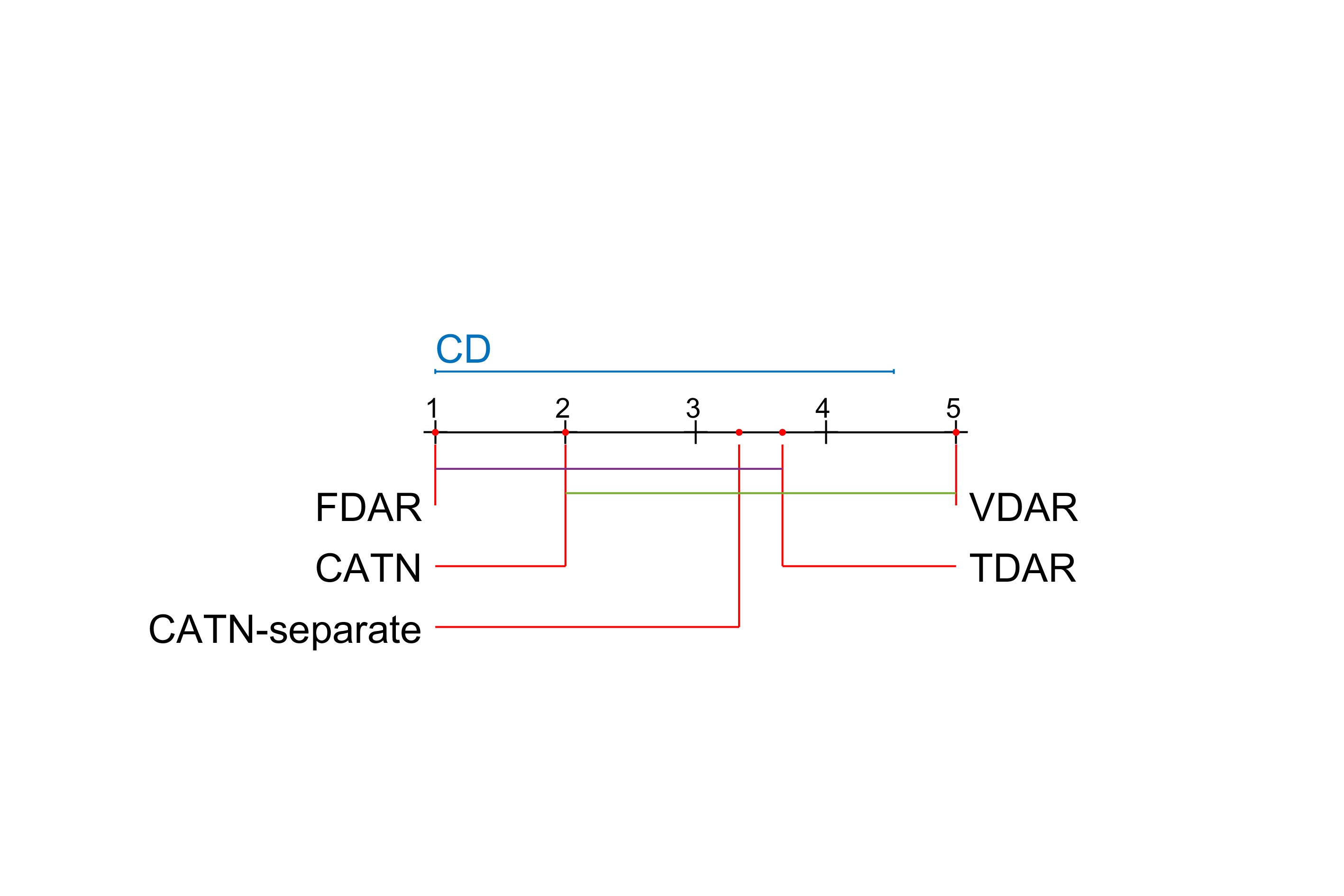}
            \caption{$NDCG@5$}
            \label{fig:ndcgCDR}
        \end{subfigure}
    \caption{Critical Difference (CD) diagrams for the comparing algorithms in SDR and CDR scenarios.}
    \label{fig:criticalDifference}
\end{figure}
\AS{This test determines whether the performance of two algorithms is significantly different by comparing their average ranks. Specifically, two algorithms are considered significantly different if their average ranks differ by at least the critical difference, which is computed as follows:}
\begin{equation}
    CD = q_\alpha \sqrt{\frac{\mathcal{K}(\mathcal{K}+1)}{6\mathcal{N}}}.
\end{equation}

\noindent \AS{Figure \ref{fig:criticalDifference} presents the CD diagrams \cite{demvsar2006statistical} for SDR and CDR scenarios over $F1@5$ and $NDCG@5$ evaluation measures. The average rank of each comparing algorithm is marked along the axis (lower ranks to the left). As seen from Figure \ref{fig:criticalDifference}, the FCF and FDAR models outperform the other algorithms in both the SDR and CDR scenarios, respectively. We observed a similar results for other values of $k$ for both evaluation metrics $F_1@k$ and $NDCG@k$.}

\begin{figure}[h!]
    \centering
    \includegraphics[width=4.5in, height=4.5in]{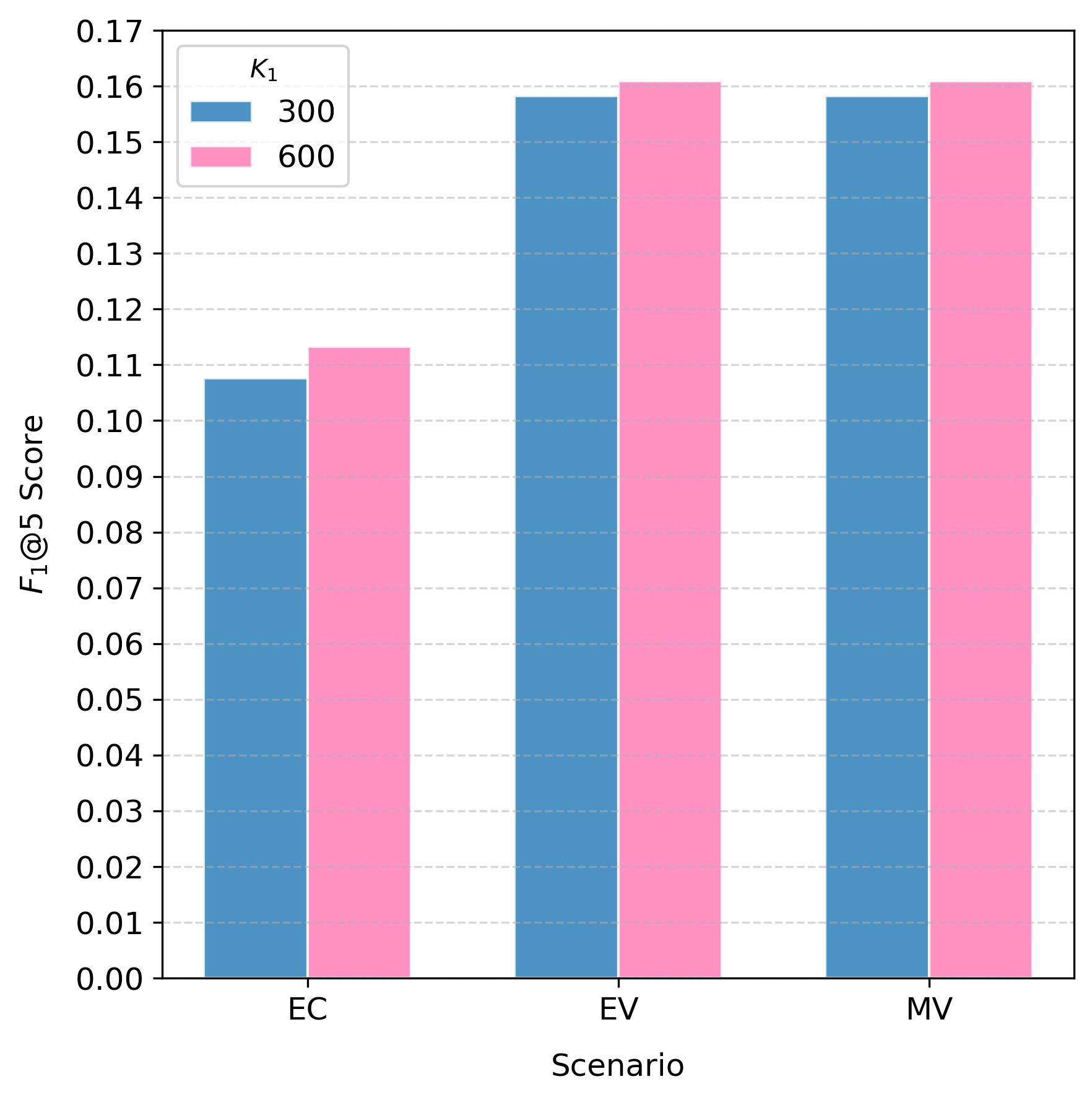}
    \caption{Comparison of $F_1@5$ Scores for FDAR Model at $K_1=300$ and $K_1=600$.}
    \label{fig:fdar_k1}
\end{figure}

\AS{In our final set of experiments, we study the impact of varying $K_1$ on the model's performance. Figure \ref{fig:fdar_k1} presents the results for the FDAR model across each scenario in terms of $F_1@5$ for $K_1=300$ and $K_1=600$. It is evident from the figure that the model performance is slightly improved for a higher value of $K_1$. This trend highlights the model's ability to encode more complex patterns, dependencies, and interactions between users and items as  $K_1$ increases.}

\AS{In summary, our proposed model seamlessly integrates visual, textual, and latent features to enhance cross-domain recommendations through domain adaptation. The experimental results consistently demonstrate its superiority over state-of-the-art methods across various evaluation metrics. These findings highlight the critical role of leveraging multimodal feature fusion strategies in cross-domain recommendation systems.}

\section{Conclusions and Future Work}
\label{sec:Conclusion}
\adamya{In this paper, a novel domain-adaptation-based CDR model is proposed that incorporates visual and textual information along with implicit feedback for preference modeling in a sparse domain.} %This paper proposed a novel domain adaptation-based cross-domain recommendation model that uses visual, textual, and implicit feedback for preference modeling in a sparse domain. %We use the domain adaptation method for a cross-domain recommendation. 
The variation in datasets in terms of entities (users and items) limits the applicability of approaches that consider overlapped entities as bridges for knowledge transfer across domains. %a different domain,  knowledge learned from the source domain is transferred to the target domain.
The domain adaptation approach focuses on the alignment of embedding of users and items for knowledge transfer. However, the distributions of embeddings across domains are generally different, making the alignment more challenging.  We combine visual and textual features with embeddings learned over implicit feedback to guide the domain adaptation process. Our experiment results corroborate the importance of multimodel domain adaptation when compared to domain adaptation approaches focused solely on text or visual information.   Extending the model to adapt all the available photos of an individual item to extract visual features is a good direction for future research. Also, making the proposed approach suitable to produce recommendations to cold-start users is another potential future work.

%In this paper, we propose a novel recommendation model which uses both visual and textual features in predicting the user preferences. We use domain adaptation method for cross-domain recommendation. Since most of the datasets suffer from sparsity problem, negative sampling creates noise in the labels. Thus knowledge learned from source domain is transferred to target domain for negative supervision while positive supervision is done from the target domain. As the distributions of both domains are different, aligning the spaces directly won't give improvement in performance. So we combine visual and textual features with embeddings (latent factors) to guide the domain adaptation process. We also did experiments by using only either of the visual and textual features to guide the alignment. But experiments show that the performance is improved when both visual and textual features (fusion of features) are used in the model. This kind of representation for both users and items is able to learn the underlying relation and thus helps in the process of recommendation. We took only one image per item to extract visual features, as a future work, we would like to extend this by considering all the available images of respective item to extract visual features for encoding user and item information. We would also like to extend this work for testing the model in regard of recommendations to cold start users.

\section*{Acknowledgements}

%Venkateswara Rao Kagita is supported by the NITW-RSM grant, NIT Warangal. 
Vikas Kumar is supported by the Start-up Research Grant (SRG), Science and Engineering Research Board, under grant number SRG/2021/001931, UGC-BSR Start-up Grant, UGC under grant number F.30-547/2021(BSR) and the Faculty Research Programme Grant, University of Delhi, under grant number IoE/2023-24/12/FRP. We would also like to thank the anonymous reviewers whose comments/suggestions helped to improve and clarify this manuscript to a large extent.

\bibliographystyle{unsrt} 
\bibliography{cas-refs}

\end{document}